\documentclass[12pt]{article}

\usepackage{amsmath,amssymb,amsfonts} \usepackage{bm} \usepackage{braket} \usepackage{graphicx}
\usepackage{xcolor}
\usepackage{float} \usepackage[numbers]{natbib} \usepackage{hyperref} \usepackage{booktabs}
\usepackage{siunitx} \usepackage[symbol]{footmisc}

  \graphicspath{{figures/}}

\makeatletter
\AtBeginDocument{%
  \let\revision@lbibitem\@lbibitem
  \def\@lbibitem[#1]#2{%
    \color{black}%
    \ifcsname revisionbib@#2\endcsname\color{red}\fi
    \ifcsname bluebib@#2\endcsname\color{blue}\fi
    \revision@lbibitem[#1]{#2}%
  }%
}

\makeatother

\begin{document}

	\begin{center}
		{\Large\bf A transversely isotropic constitutive model for semisoft elasticity in main-chain liquid crystal elastomers}\par\vspace{6pt}
		Donghao Li, Xin Yi and Fan Feng\footnote[1]{Corresponding author: fanfeng@pku.edu.cn}\par
		\small School of Mechanics and Engineering Science, Peking University, Beijing 100871, China
	\end{center}

\begin{abstract}
Monodomain main-chain liquid crystal elastomers exhibit pronounced initial stiffness anisotropy and broad semisoft plateaus that can follow a substantial onset strain. However, the standard neo-classical model at fixed chain anisotropy predicts equal principal Young's moduli before director reorientation, while its transversely isotropic extension with a constant semisoft modulus constrains the relation between rotation onset and subsequent stress growth. Here, we formulate a finite-strain, isothermal elastic model combining transversely isotropic network elasticity with a deformation-dependent quadratic semisoft penalty. At fixed mean anisotropy, a reduced statistical closure assumes linear contraction of the deformation-sensitive component of effective local chain-anisotropy fluctuations. Together with probability conservation and a weak fluctuation approximation, this assumption yields an exponential transition from an initial to a residual semisoft modulus. The network energy retains the classical ideal soft modes, and the formulation recovers the de Gennes' strain--director coupling model in the infinitesimal limit. Analytical relations are derived for the symmetry-preserving parallel branch and for the onset and stress response of continuous reorientation under perpendicular tension. The rotating-stage stress depends on both the current semisoft modulus and its derivative with respect to the first deformation invariant, allowing delayed rotation together with a substantially flatter response than in the constant-modulus limit. Coupled director rotation, shear and lateral contraction are computed under oblique tension. Comparisons with orientation-dependent tensile measurements using a common parameter set capture the principal stiffness contrast and broad perpendicular plateau, with discrepancies remaining in oblique stresses and intermediate-angle initial moduli. The framework connects measurable tensile features to network anisotropy and finite-strain reorientation resistance within a homogeneous constitutive description, paving the way for designing soft actuators, compliant structures, and related applications.
\end{abstract}



\section{Introduction}
\label{sec:introduction}

Liquid crystal elastomers (LCEs) combine the orientational order of liquid crystal mesogens with the large, reversible deformability of cross-linked polymer networks \citep{warner2003}. Their ability to convert changes in molecular order into macroscopic deformation makes them promising materials for intelligent actuation and programmable soft structures. Representative demonstrations include muscle-like thermomechanical actuation \citep{thomsen2001muscle} and spatially programmed shape transformations \citep{ware2015voxelated}. Monodomain main-chain LCEs provide a demanding test of constitutive models because pronounced directional stiffness can coexist with delayed reorientation and a broad, nearly flat plateau under perpendicular tension \citep{okamoto2021effect}. Their response requires a joint description of network anisotropy and resistance to director reorientation.

Tensile measurements on monodomain main-chain LCEs reveal pronounced differences between the initial moduli parallel and perpendicular to the director, together with substantial variations at intermediate loading angles \citep{merkel2018thermomechanical,ware2016localized,okamoto2021effect}. Complementary measurements on an all-acrylate LCE also demonstrate angular variations in both the initial stiffness and the subsequent nonlinear response \citep{mistry2019mechanical}. Under perpendicular tension, some main-chain networks exhibit a pronounced reduction in tangent stiffness after a finite onset strain, followed by a broad, nearly constant nominal-stress plateau. In the material studied by \citet{okamoto2021effect}, this plateau narrows as the loading direction approaches the initial director and becomes indistinct at smaller angles. The same study distinguished oppositely rotating director populations under perpendicular tension from uniform rotation toward the initially biased direction under oblique tension. Simultaneous measurements of stress, director orientation and strain further show that reorientation under tilted stretching is coupled to shear and lateral deformation \citep{wei2023stressorder}. The magnitude of the stiffness contrast and the form of the soft response also depend on network composition and preparation history, including the alignment imposed during polymerization \citep{merkel2018thermomechanical,wei2025synthesis}. Together, these observations require a model to account for directional stiffness and the finite onset and broad soft response under perpendicular tension, while predicting the coupled oblique response for the same material.

The neo-classical theory couples the statistics of an anisotropic Gaussian-chain network to the nematic director, providing a molecular basis for soft elasticity \citep{bladon1993transitions,warner1994soft}. Its ideal soft modes combine stretch, shear and director rotation without additional network energy \citep{olmsted1994rotational}. The classical semisoft correction introduces a finite energetic resistance to reorientation and lifts the degeneracy of these ideal states \citep{verwey1997compositional,biggins2008semisoft}. These descriptions have explained rotation thresholds, the accompanying shear, and stripe-domain formation under constrained loading \citep{verwey1996elastic,conti2002semisoft}. To describe time-dependent deformation and director reorientation, \citet{zhang2019continuum} formulated LCEs as dissipative ordered solids and derived coupled governing equations applicable to large deformation and large director rotation. This framework has been applied to stripe domain formation \citep{zhang2020nucleation} and director-mediated redistribution of stress and strain around a circular hole \citep{jiang2021stress}. 

A quantitative limitation of the standard neo-classical model at fixed chain anisotropy $r$ is that it predicts equal initial Young's moduli under parallel and perpendicular tension before director reorientation, despite the substantial modulus contrast measured experimentally \citep{li2025extended}. Early studies addressed this discrepancy by coupling network deformation to changes in nematic order \citep{pereira2001mechanical,jin2010thermomechanical}. In particular, analytical calculations showed how order relaxation and strain-induced biaxiality produce unequal principal moduli and their temperature dependence \citep{zeng2010anisotropic}. More recent models have combined thermal and mechanical changes in nematic order with director rotation and additional anisotropic network elasticity \citep{kang2025thermomechanical}. The micro-stretch theory of \citet{jiang2026microstretch} further incorporates viscous mesogen ordering, network relaxation, and spatial gradients of the director and order parameters. Along a different route, \citet{li2025extended} introduced transversely isotropic network elasticity into the neo-classical construction, allowing unequal principal Young's moduli while retaining the ideal soft modes and their semisoft regularization. From a statistical network perspective, \citet{lamont2024nonaffine} described polymer-chain and mesogen motions separately and allowed non-affine chain deformation, capturing stiffness anisotropy and angle-dependent softening. These developments establish several routes to describing stiffness anisotropy together with director-mediated deformation, based on order evolution, transversely isotropic network elasticity or non-affine chain motion.

Beyond directional stiffness, the finite-strain tensile response also depends on the constitutive form of the semisoft resistance. The classical quadratic correction relates this resistance to compositional fluctuations and conventionally represents their effect through a constant coefficient \citep{verwey1997compositional,biggins2008semisoft}. Extensions have modified either the nonlinear penalty or its coefficient. \citet{xu2021electro} introduced a higher-order semisoft penalty to change the electrically induced solid Freedericksz transition from subcritical to supercritical, demonstrating that the nonlinear form of the resistance can alter the character of director reorientation. For mechanical loading, \citet{zhou2025modified} adopted a semisoft coefficient that decreases with deformation to improve the description of the initial high stiffness and subsequent low-slope tensile response, and extended the formulation to viscoelastic and nonuniform deformation. In the transversely isotropic model of \citet{li2025extended}, the homogeneous perpendicular rotating branch has a nominal-stress slope proportional to the constant semisoft modulus. The same modulus enters the onset condition, so that the onset and subsequent stress growth are linked when the network parameters and $r$ are fixed. This constitutive connection restricts the simultaneous representation of a substantial onset strain and an extended, nearly flat plateau.

Taken together, experiments pose a joint constitutive requirement: a description of the directional initial stiffness must also accommodate a finite reorientation threshold and a broad, nearly flat plateau under perpendicular tension. The central issue is how the rotation onset constrains the subsequent stress increase once the network stiffness has been specified. Making this relation explicit connects observable features of the tensile response to the contributions of network anisotropy and evolving semisoft resistance. Using a common parameter set across loading directions then tests the consistency of this physical description and helps identify the limits of the reduced elastic model.

Building on the construction of transversely isotropic soft elasticity in \citet{li2025extended} and the deformation-dependent semisoft approach of \citet{zhou2025modified}, we combine a transversely isotropic Saint Venant--Kirchhoff network energy with a variable quadratic semisoft penalty. The selected network energy preserves the ideal soft modes, with $C_1$ and $C_4$ determining the principal tensile stiffnesses and $C_5$ controlling network shear in planes containing the reference director. Within the weak fluctuation approximation to compositional semisoftness \citep{verwey1997compositional}, we adopt a reduced statistical closure at fixed mean $r$. The effective local chain-anisotropy fluctuations are decomposed into uncorrelated residual and deformation-sensitive components, and the latter is assumed to obey a linear contraction law. Probability conservation then yields an exponential transition from the initial to the residual semisoft modulus. We derive the initial tensile and shear moduli and analytical relations for the onset and subsequent stress response under perpendicular tension. These relations identify how the current semisoft modulus enters the onset condition and how both the modulus and its derivative with respect to the deformation invariant $I_1$ contribute to the rotating-stage stress. These results quantify how deformation-dependent semisoft resistance modifies the constant-modulus connection between rotation onset and subsequent stress growth. We then compute nominal stress, director orientation, shear and lateral contraction for different initial loading angles. The predicted tensile curves and initial moduli are compared with experimental measurements using one common material parameter set across loading orientations.

Our work focuses on the accurate constitutive modeling of main-chain LCE to capture both the anisotropic initial modulus and the large-strain plateau.
The paper is organized as follows. Section~\ref{sec:constitutive_model} establishes the continuum balance laws and general constitutive relations. Section~\ref{sec:soft_semisoft_elasticity} specifies the network energy and semisoft closure and derives the linearized elastic moduli. Section~\ref{sec:uniaxial_deformation} analyzes parallel, perpendicular and oblique tension, including the principal branches and coupled director, shear and lateral responses. Section~\ref{sec:material_control} examines the effects of material parameters and compares the predictions with orientation-dependent tensile measurements. Section~\ref{sec:conclusions} summarizes the main findings.

\section{Continuum framework}
\label{sec:constitutive_model}

We consider the isothermal, quasistatic response of an incompressible monodomain liquid crystal elastomer. The deformation and director orientation are treated as independent state variables. Frank energy, dissipation, and external body forces and couples are omitted. The resulting local formulation supplies the equilibrium equations for the homogeneous tensile analysis. The balance laws and their relation to objectivity are derived in Appendix~\ref{app:continuum_details}.

\subsection{Kinematics}
\label{subsubsec:kinematics}

A continuum body of nematic liquid crystal elastomers (LCEs) is modeled as a three-dimensional differentiable material manifold $\mathcal{B}$. Its motion is described by a time-dependent family of smooth embeddings $\varphi(t):\mathcal{B}\rightarrow\mathbb{R}^3$. The reference configuration at time $t_0$ and the current configuration at time $t$ are defined as $\mathcal{B}_0:=\varphi(t_0)(\mathcal{B})$ and $\mathcal{B}_t:=\varphi(t)(\mathcal{B})$, respectively. To characterize the local average orientation of the liquid crystal mesogens, each material point $\mathbf{x}\in\mathcal{B}_0$ is endowed with a unit director field $\mathbf{n}_0:\mathcal{B}_0\rightarrow\mathbb{S}^2$, where $\mathbb{S}^2$ denotes the two-dimensional unit sphere embedded in $\mathbb{R}^3$. The deformation from $\mathcal{B}_0$ to $\mathcal{B}_t$ is defined by $\varphi_t:=\varphi(t)\circ\varphi(t_0)^{-1}$, and the current position of a reference material point $\mathbf{x}$ is $\mathbf{y}:=\varphi_t(\mathbf{x})\in\mathcal{B}_t$. During deformation, the director may also evolve, and its current orientation is described by $\mathbf{n}:\mathcal{B}_t\rightarrow\mathbb{S}^2$.

At the macroscopic scale, we adopt a local constitutive assumption in which the elastic energy density per unit reference volume depends on the deformation gradient $\mathbf F:=\nabla_{\mathbf x}\mathbf y$ and the current director $\mathbf n$, with the reference director $\mathbf n_0$ characterizing the material anisotropy
\begin{equation}
\label{eq:general_free_energy}
W=\hat W
\left(
\mathbf{F},\mathbf{n};\mathbf{n}_0
\right).
\end{equation}
Here, $\nabla_{\mathbf x}(\cdot)$ denotes the right gradient with respect to $\mathbf x\in\mathcal B_0$. The contribution of director gradient $\nabla_{\mathbf x}\mathbf n$ associated with Frank elasticity is neglected, assuming that the characteristic length scale of director variations is much larger than the nematic penetration length \cite{warner2003}.

\subsection{Balance laws and state equations}
\label{subsec:model_structure}

For a reference configuration $\mathcal B_0$ of LCEs, the momentum balance equation yields
\begin{equation}
\label{eq:momentum_baance}
\mathrm{Div} \mathbf P=\mathbf 0, \, \rm{in} \, \mathcal{B}_0 \qquad  \mathbf{P}\cdot{\mathbf N} = \mathbf{P}_{\mathbf N}, \, \rm{on} \, \partial \mathcal{B}_0
\end{equation}
Here, $\mathbf P$ is the first Piola--Kirchhoff stress, $\mathrm{Div}$ denotes the right divergence operator, $\mathbf P_{\mathbf N}$ is the traction on the surface $\partial \mathcal{B}_0$ with normal vector $\mathbf N$ . 

The first Piola--Kirchhoff stress can be determined by the elastic energy (Eq. (\ref{eq:general_free_energy}))
\begin{equation}
\label{ch2:eq:Piola_stress}
\mathbf P = -p\mathbf F^{-\mathrm T}
+\frac{\partial W}{\partial\mathbf F},
\end{equation}
by which the Cauchy stress can be derived as
\begin{equation}
\label{eq:Cauchy_stress_general}
\mathbf t =  \mathbf P \mathbf F^{\rm T} = -p\mathbf I + \frac{\partial W}{\partial\mathbf F}\mathbf F^{\rm T}.
\end{equation}
 The multiplier $p$ enforces the volume preserving constraint $J:=\rm{det}\mathbf F=1$.
 
Neglecting the effect of director gradient, the local micro-momentum balance can be simplified as (detailed in Appendix \ref{app:continuum_details})
\begin{equation}
\label{ch2:eq:director_equilibrium}
\mathbf g=\mathbf0, \, \rm{in} \, \mathcal{B}_0
\end{equation}
where $\mathbf g:= \gamma\mathbf n-{\partial W}/{\partial\mathbf n}$ is the intrinsic body force associated with director rotation. The multiplier $\gamma$ enforces the unit-length constraint $M:=\mathbf n\cdot \mathbf n=1$.

\subsection{Director equilibrium and angular momentum balance}
\label{subsubsec:angular_momentum}
Objectivity requires the energy to remain invariant under an arbitrary infinitesimal rigid-body rotation,
\begin{equation}
\label{ch2:eq:objectivity}
\mathbf y^{*}
=
\mathbf y
+
\zeta\boldsymbol{\Omega}^{\rm skew}\mathbf y,
\qquad
\mathbf n^{*}
=
\mathbf n
+
\zeta\boldsymbol{\Omega}^{\rm skew}\mathbf n,
\end{equation}
where $\zeta$ is an infinitesimal parameter and $\boldsymbol{\Omega}^{\rm skew}$ is an arbitrary skew-symmetric tensor. Differentiating the energy at $\zeta=0$ gives
\begin{equation}
\label{eq:objectivity_identity}
\begin{aligned}
0=
\left.
\frac{\partial}{\partial\zeta}
W(\mathbf F^{*},\mathbf n^{*};\mathbf n_0)
\right|_{\zeta=0} =
\left(
\frac{\partial W}{\partial\mathbf F}\mathbf F^{\rm T}
+
\frac{\partial W}{\partial\mathbf n}\otimes\mathbf n
\right)
:
\boldsymbol{\Omega}^{\rm skew}.
\end{aligned}
\end{equation}
Consequently,
\begin{equation}
\label{eq:Ericksen_identity}
\boldsymbol{\epsilon}:
\left(
\frac{\partial W}{\partial\mathbf F}\mathbf F^{\rm T}
+
\frac{\partial W}{\partial\mathbf n}\otimes\mathbf n
\right)
=
\mathbf0,
\end{equation}
where $\boldsymbol{\epsilon}$ is the third-order permutation tensor. It is noticed that, in Eq. (\ref{eq:Ericksen_identity}), since the director balance $\partial W/\partial\mathbf n=\gamma\mathbf n$ eliminates the second term, and the hydrostatic pressure $-p \mathbf I$ makes no contribution to the contraction, the angular momentum balance reduces to
\begin{equation}
\label{eq:angular_momentum_balance}
\boldsymbol{\epsilon}:\mathbf t=\mathbf0,
\end{equation}
which is equivalent to $\mathbf t=\mathbf t^{\rm T}$. Conversely, Eqs.~\eqref{eq:Ericksen_identity} and \eqref{eq:angular_momentum_balance} imply that $\mathbf n \times\partial W/\partial\mathbf n=\mathbf0$, recovering the director balance. Thus, the angular momentum balance ($\boldsymbol{\epsilon}:\mathbf t=\mathbf0$) and director rotation balance ($\boldsymbol{g}=\boldsymbol{0}$) are equivalent by means of objectivity. The corresponding balance relation with director gradient contributions is derived in Appendix~\ref{app:continuum_angular_momentum}.

\section{Elastic energy}
\label{sec:soft_semisoft_elasticity}

\subsection{Constituents of elastic energy}
\label{subsec:energy_constituents}

The elastic energy per unit reference volume is written as
\begin{equation}
\label{eq:total_energy}
W = W_{\rm TI}^{\rm soft} + W^{\rm semi}.
\end{equation}
The first term $W_{\rm TI}^{\rm soft}:=\hat W_{\rm TI}^{\rm soft}
(\mathbf E^{\rm e},\mathbf n_0)$ characterize the soft elasticity of transversely isotropic network with material axis $\mathbf n_0$. The second term $W^{\rm semi}=\hat{W}^{\rm semi}(\mathbf F, \mathbf n; \mathbf n_0)$, penalizes director reorientation relative to the network. 

Here, 
\begin{equation}
\label{eq:elastic_Green_strain}
\mathbf E^{\rm e}
=
\frac12
\left[
(\mathbf F^{\rm e})^{\rm T}\mathbf F^{\rm e}
-
\mathbf I
\right]
\end{equation}
is the Green--Lagrange strain determined by  the elastic deformation decomposition
\begin{equation}
\label{eq:elastic_deformation_gradient}
\mathbf F^{\rm e}
=
(\mathbf V^{\rm s})^{-1}
\mathbf F
\mathbf U^{\rm s},
\end{equation}
with the current and reference spontaneous stretch tensors defined by
\begin{equation}
\label{eq:spontaneous_stretch}
\begin{aligned}
\mathbf V^{\rm s}
&=
r^{-1/6}
\left[
\mathbf I
+
\left(r^{1/2}-1\right)
\mathbf n\otimes\mathbf n
\right],
\\
\mathbf U^{\rm s}
&=
r^{-1/6}
\left[
\mathbf I
+
\left(r^{1/2}-1\right)
\mathbf n_0\otimes\mathbf n_0
\right],
\end{aligned}
\end{equation}
where parameter $r$ is the step-length ratio of polymer conformation. 

The energy decomposition in Eq.~\eqref{eq:total_energy} separates ordinary network elasticity from resistance to director reorientation. In the next two subsections, we specify these contributions and establish their infinitesimal response before considering finite tensile deformations.

\subsection{Transversely isotropic network energy and ideal soft modes}
\label{subsec:soft_elasticity}

The widely utilized neo-classical soft elastic energy\cite{bladon1993transitions} can be written as a function of elastic Green--Lagrange strain
\begin{equation}
\label{eq:neo_classical_energy}
\begin{aligned}
W_{\rm NC}^{\rm soft}=
\mu
{\rm tr}(\mathbf E^{\rm e})
\end{aligned}
\end{equation}
Its elastic stiffness is controlled by a single modulus $\mu$. To allow transverse isotropy in the network response, and following the method in \citep{li2025extended}, we generalize the soft elastic energy as a transversely isotropic Saint Venant--Kirchhoff form
\begin{equation}
\label{eq:TI_soft_energy}
W_{\rm TI}^{\rm soft}
=
\frac12
\mathbf E^{\rm e}
:
{\mathbb C}^{\rm TI}
:
\mathbf E^{\rm e}.
\end{equation}
Here, the fourth-order tensor $\mathbb C^{\rm TI}$ represents a general stiffness tensor that is invariant under rotations about the director $\boldsymbol{n}_0$. Based on the tensor representation theory (\cite{walpole1984fourth,lubarda2008elastic}), an incompressible network carries three independent components $C_1$, $C_4$, and $C_5$ in the Voigt notation (Appendix~\ref{appB:director_relaxation}). In this case, we have
\begin{equation}
\label{eq:TI_stiffness}
\begin{aligned}
{\mathbb C}^{\rm TI}
:={}&
4C_4\mathbb I^{\rm s}
+4(C_5-C_4)\mathbb A_0+
 2(C_1+2C_4-4C_5)
\mathbf N_0\otimes\mathbf N_0.
\end{aligned}
\end{equation}
with $\mathbf N_0=\mathbf n_0 \otimes \mathbf n_0$, $\mathbb{I}^{\rm s}_{ijkl}=\left( \delta_{ik}\delta_{jl}+\delta_{il}\delta_{jk} \right)/2$ and $\mathbb{A}_{0,ijkl}=(N_{0,ik}\delta_{jl}+N_{0,il}\delta_{jk}+N_{0,jk}\delta_{il}+N_{0,jl}\delta_{ik})/2$. The positive definiteness of soft elastic energy requires $C_1$, $C_4$, $C_5>0$. The projection form derived in Appendix~\ref{appB:quadratic_coupling} shows that the network energy reaches its minimum value only if $\mathbf E^{\rm e}=\mathbf0$, which is equivalent to $\mathbf F^{\rm e} \in \mathrm{SO}(3)$. Soft deformation
\begin{equation}
\label{eq:ideal_soft_mode}
\mathbf F^{\rm soft}
=
\mathbf V^{\rm s}\mathbf O(\mathbf U^{\rm s})^{-1},
\qquad
\mathbf O\in{\rm SO}(3)
\end{equation}
are therfore obtained, which means that director rotation changes the configuration without elastic network distortion \citep{olmsted1994rotational}. 

\subsection{Deformation-dependent semisoft energy and statistical closure}
\label{subsec:semisoft_elasticity}

The ideal soft modes in Eq.~\eqref{eq:ideal_soft_mode} accommodate director rotation without storing network elastic energy. A semisoft contribution is introduced to represent the finite resistance to this motion. With a constant semisoft modulus, the onset of perpendicular reorientation and the subsequent stress growth are linked when the network parameters and $r$ are fixed \citep{li2025extended}. This connection restricts the description of main-chain LCEs that exhibit a substantial onset strain followed by a broad, nearly flat plateau \citep{okamoto2021effect}. We therefore allow the semisoft resistance to depend reversibly on deformation while retaining the classical quadratic form \citep{biggins2008semisoft}:
\begin{equation}
\label{eq:semisoft}
W^{\rm semi}
=
D_{\rm ss}
\left|\mathbf n^{\rm F}\right|^2,
\end{equation}
with
\begin{equation}
\label{eq:semisoft_distortion}
\mathbf n^{\rm F}
:=
\mathbf F^{\rm T}\mathbf n
-
\left(
\mathbf F^{\rm T}\mathbf n\cdot\mathbf n_0
\right)\mathbf n_0.
\end{equation}
The vector $\mathbf n^{\rm F}$ is the component of $\mathbf F^{\rm T}\mathbf n$ perpendicular to $\mathbf n_0$, and $D_{\rm ss}$ measures the energetic resistance associated with this distortion. To specify its deformation dependence at fixed mean chain anisotropy $r$, we adopt a reduced statistical closure for effective local chain-anisotropy fluctuations. The starting point is the compositional-fluctuation interpretation of semisoftness \citep{verwey1997compositional,warner2003}
\begin{equation}
\label{eq:SI_Warner_relation}
D_{\rm ss}
\propto
\left\langle\frac1{r^{(\nu)}}\right\rangle_\nu-\frac1r,
\qquad
r=\langle r^{(\nu)}\rangle_\nu.
\end{equation}
Here $\nu$ labels a network strand in a representative local ensemble, and $\langle\cdot\rangle_\nu$ denotes the ensemble average. In the reduced description adopted below, $r^{(\nu)}$ represents an effective local chain anisotropy evaluated at the current deformation, while its mean $r$ is prescribed. The quantity $\delta r^{(\nu)}$ denotes a statistical deviation from this mean. Writing $\delta r^{(\nu)}=r^{(\nu)}-r$, weak fluctuations about the fixed mean give
\begin{equation}
\label{eq:SI_variance_relation}
\left\langle\frac1{r^{(\nu)}}\right\rangle_\nu-\frac1r
\approx\frac{\Delta}{r^3},
\qquad
\Delta:=\left\langle(\delta r^{(\nu)})^2\right\rangle_\nu.
\end{equation}
We interpret $\Delta$ as an effective variance conditioned on the current deformation. For incompressible states, we choose $I_1={\rm tr}(\mathbf F^{\rm T}\mathbf F)\ge3$ and $\zeta:=I_1-3$ as an objective scalar measure of deformation. Restricting this additional constitutive dependence to $I_1$ is a reduced modeling assumption.

To distinguish persistent and deformation-sensitive contributions, we assume
$\delta r^{(\nu)}=\delta r_\infty^{(\nu)}+a^{(\nu)}(\zeta)$.
The fixed residual component $\delta r_\infty^{(\nu)}$ represents effective heterogeneity associated with the frozen network structure that remains active in the semisoft response. The component $a^{(\nu)}$ represents an additional effective contrast that is assumed to weaken as local chain--mesogen configurations adjust with deformation. Both components are taken to have zero mean and to be uncorrelated, so that their contributions to the variance are additive. This decomposition describes effective anisotropy fluctuations without changing the permanent network architecture. 

To describe contraction toward $a=0$, we retain the leading linear term of a smooth restoring transport field and take its coefficient to be constant
\begin{equation}
\label{eq:SI_transport_velocity}
\mathcal J
:=
\frac{{\rm d}a}{{\rm d}\zeta}
=-\frac m2a,
\qquad m\ge0.
\end{equation}
Here $\mathcal J$ is the transport velocity in fluctuation space. This phenomenological assumption is motivated by a leading-order expansion near a stable relaxed state, with the relaxation coefficient taken as constant over the deformation range considered; the resulting linear restoring term is mathematically analogous to the deterministic drift in the Ornstein–Uhlenbeck process \cite{uhlenbeck1930theory}. Linear transport preserves the zero mean and closes the variance equation without prescribing the shape of the initial distribution. The factor $m$ is the decay coefficient of the deformation-sensitive variance. Since $\zeta$ labels the current deformation, this relation describes a reversible state dependence of effective fluctuations. Together with probability conservation, the assumed contraction gives
\begin{equation}
\label{eq:SI_variance_solution}
\Delta(I_1)
=\Delta_\infty+
(\Delta_0-\Delta_\infty)\exp[-m(I_1-3)].
\end{equation}
The probability transport and variance calculation are given in
Appendix~\ref{appB:statistical_closure}.

At fixed $r$, the variance relation motivates the constitutive law
\begin{equation}
\label{eq:semisoft_law}
D_{\rm ss}(I_1)
=
D_{\rm ss}^{\infty}
+
\left(D_{\rm ss}^{0}-D_{\rm ss}^{\infty}\right)
\exp[-m(I_1-3)],
\end{equation}
with
\begin{equation}
\label{ch3:eq:semisoft_parameter_domain}
D_{\rm ss}^{0}\ge D_{\rm ss}^{\infty}\ge0,
\qquad m\ge0.
\end{equation}
The parameter $m$ sets the deformation scale of the transition, and the constant-modulus limit is obtained when $m=0$ or $D_{\rm ss}^{0}=D_{\rm ss}^{\infty}$. 

The resulting coefficient $D_{ss}$ is a state function of $I_1$ at fixed mean anisotropy. It describes reversible changes in effective fluctuations, without assigning a time scale to $m$ or changing the permanent network architecture. This dependence on the current deformation must also be retained when differentiating the energy to obtain stress. Consequently, the semisoft stress involves both $D_{\rm ss}$ and its derivative with respect to $I_1$, as shown in the following subsection.

\subsection{Specific formula of Cauchy stress}

The energies specified above determine the Cauchy stress required to impose the mechanical boundary conditions and calculate the tensile response. Substituting Eqs.~\eqref{eq:total_energy}, \eqref{eq:TI_soft_energy} and \eqref{eq:semisoft} into Eq.~\eqref{eq:Cauchy_stress_general} gives
\begin{equation}
\label{eq:stress_tensor}
\begin{aligned}
\mathbf t =-p\mathbf I
+\mathbf t^{\rm soft}
+\mathbf t^{\rm semi},
\end{aligned}
\end{equation}
where the soft elastic stress 
\begin{equation}
\label{eq:soft_stress}
\mathbf t^{\rm soft} := \frac{\partial W_{\rm TI}^{\rm soft}} {\partial\mathbf F} \mathbf F^{\rm T} = (\mathbf V^{\rm s})^{-1}\mathbf t^{\rm e}\mathbf V^{\rm s}
\end{equation}
is determined by the pure elastic Cauchy stress
\begin{equation}
\label{eq:elastic_stress}
\mathbf t^{\rm e} := \frac{\partial W_{\rm TI}^{\rm soft}}
{\partial\mathbf F^{\rm e}}
(\mathbf F^{\rm e})^{\rm T}
= \mathbf F^{\rm e} (\mathbb C^{\rm TI}:\mathbf E^{\rm e})(\mathbf F^{\rm e})^{\rm T},
\end{equation}
which is introduced to characterize the effect of network elasticity. The semisoft term 
\begin{equation}
\label{eq:semisoft_stress}
\begin{aligned}
\mathbf t^{\rm semi}
&:=\frac{\partial W^{\rm semi}}{\partial\mathbf F}\mathbf F^{\rm T}=2D_{\rm ss}\mathbf n\otimes\mathbf F\mathbf n^{\rm F}+2D_{\rm ss}'|\mathbf n^{\rm F}|^2\mathbf F\mathbf F^{\rm T}
\end{aligned}
\end{equation}
is given by the derivative of the semisoft energy.
Here $D_{\rm ss}':=\partial D_{\rm ss}/\partial I_1$. The first term arises from differentiation of the quadratic distortion measure $|\mathbf n^{\rm F}|^2$, and the second from the deformation dependence of its coefficient $D_{\rm ss}$. The latter vanishes in the constant-modulus limit.

The infinitesimal response is examined next to relate the material parameters to the initial tensile and shear moduli.

\subsection{Linearized elasticity and director-relaxed moduli}
\label{subsec:linearized_elasticity}

To determine the initial directional stiffness, we linearize the energy about $\mathbf F=\mathbf I$ and $\mathbf n=\mathbf n_0$, obtaining a de Gennes-type quadratic energy \citep{de1980weak}. For the homogeneous equilibrium response considered here, the director is allowed to equilibrate with the imposed infinitesimal deformation. The relevant moduli therefore follow from minimizing this quadratic energy over the relative director rotation (as detailed in Appendix \ref{appB:quadratic_coupling}). This relaxation defines the effective shear coefficient $C_5^{\rm eff}$. The expansion and relaxation are detailed in Appendix~\ref{app:linear_compatibility}, and the resulting director-relaxed engineering moduli are
\begin{equation}
\label{eq:SI_apparent_moduli}
\begin{aligned}
E_\parallel
&=2(C_1+C_4),
&
E_\perp
&=\frac{8C_4(C_1+C_4)}{C_1+2C_4},
\\
G_{\rm d}
&=2C_5^{\rm eff}
=\frac{2rC_5D_{\rm ss}^{0}}
{{C_5(r-1)^2}/{r}+D_{\rm ss}^{0}},
&
G&=2C_4.
\end{aligned}
\end{equation}
Here $G_{\rm d}$ and $G$ are the longitudinal and transverse shear
moduli. $E_{\parallel}$ and $E_{\perp}$ are the longitudinal and transverse Young's moduli. Incompressibility implies $E_\perp=4E_\parallel G/(E_\parallel+G)$, so only three of these four moduli are independent.

For uniaxial stress at initial angle $\theta_0$ to $\mathbf n_0$, the effective (director-relaxed) Young's modulus is
\begin{equation}
\label{ch3:eq:Etheta}
\frac1{E_\theta}
=
\frac{\cos^4\theta_0}{E_\parallel}
+
\frac{\sin^4\theta_0}{E_\perp}
+
\left(
\frac1{G_{\rm d}}-\frac1{E_\parallel}
\right)\sin^2\theta_0\cos^2\theta_0.
\end{equation}
Equation~\eqref{eq:SI_apparent_moduli} shows that $C_1$ and $C_4$ determine the principal Young's moduli. The joint contribution of $C_5$ and $D_{\rm ss}^{0}$ enters through the relaxed longitudinal shear modulus $G_{\rm d}$ and therefore affects the oblique modulus $E_\theta$.
For fixed $r$ and $D_{\rm ss}^{0}>0$,
\begin{equation}
\label{ch3:eq:shear_modulus_bound}
0<G_{\rm d}<
\frac{2r^2D_{\rm ss}^{0}}{(r-1)^2},
\end{equation}
with the upper limit approached as $C_5\rightarrow\infty$. Angular-modulus measurements therefore constrain a joint combination of $C_5$ and $D_{\rm ss}^{0}$; separating them requires additional information.

The reduced choice $C_1=2C_4$, $C_5=C_4$, and $\mu=2C_4$ matches the neo-classical quadratic response at the same $r$ and $D_{\rm ss}^{0}$. Their finite-strain network energies remain different. Since $I_1-3$ is of second order near the reference state, $D_{\rm ss}^{\infty}$ and $m$ do not enter the initial quadratic energy. Thus, at fixed $r$, $D_{\rm ss}^{0}$ and network parameters, $D_{\rm ss}^{\infty}$ and $m$ can modify finite-strain reorientation without changing the initial moduli. The tensile analysis below examines how this distinction appears in the reorientation onset and subsequent stress response.


\section{Uniaxial tension: principal branches and orientation effects}
\label{sec:uniaxial_deformation}

In this section, we examine how the initial director orientation affects the finite-strain tensile response through the coupling between network distortion and director reorientation. The analysis considers isothermal uniaxial tension at fixed mean chain anisotropy $r$, with shear unconstrained. Parallel loading admits a branch without director rotation, while perpendicular loading can produce two equal-energy variants with opposite director rotations and opposite shear. These principal loading cases provide the reference for examining how oblique loading breaks this symmetry and changes the tensile stress, shear and lateral contraction.

\subsection{Homogeneous kinematics and governing equations}
\label{ch3:subsec:reduction}

Under homogeneous uniaxial stress loading,  the stress boundary conditions are expressed as
\begin{equation}
\label{eq:stress_boundary}
t_{11}=\sigma,\qquad
t_{22}=t_{33}=0,\qquad t_{12}=t_{21}=0.
\end{equation}
In this case, the deformation can be prescribed as a coupling of stretch with shear. Let the Cartesian coordinate base $\mathbf e_1$  along the loading direction and another base $\mathbf e_2$ in the director rotation plane, s.t. $\mathbf e_1 \perp \mathbf e_2$. In this coordinate, the deformation gradient is expressed as
\begin{equation}
\label{ch3:eq:kinematics}
[\mathbf F]=
\begin{bmatrix}
\lambda_1&\lambda_{12}&0\\
0&\lambda_2&0\\
0&0&\lambda_3
\end{bmatrix},
\qquad
\lambda_3=(\lambda_1\lambda_2)^{-1},
\end{equation}
where $\lambda_1\ge1$. The directors in reference and current configuration can be respectively denoted by
\begin{equation}
\label{ch3:eq:kinematics_director}
\mathbf n_0=\cos\theta_0\,\mathbf e_1+\sin\theta_0\,\mathbf e_2,
\qquad
\mathbf n=\cos\theta\,\mathbf e_1+\sin\theta\,\mathbf e_2.
\end{equation}

By eliminating the hydrostatic pressure $p$ with $t_{33}=0$, the uniaxial stress condition of Eq. (\ref{eq:stress_boundary}) and angular-momentum balance of Eq. (\ref{eq:angular_momentum_balance}) give the governing equations
\begin{equation}
\label{ch3:eq:governing_equations}
t_{11}-t_{22}=\sigma,\qquad
t_{22}-t_{33}=0,\qquad
t_{12}=0,\qquad t_{12}-t_{21}=0.
\end{equation}
The last three equations determine $\lambda_2,\lambda_{12}$ and $\theta$ at prescribed $\lambda_1$, the first equation determines the nominal tensile stress is $P_{11}=\sigma/\lambda_1$.

\subsection{Parallel tension}
\label{ch3:subsec:parallel}

For $\theta_0=0$, the governing equations admit the branch $\theta=\lambda_{12}=0$. The spontaneous stretch tensors then coincide and commute with the diagonal deformation, so that $\mathbf F^{\rm e}=\mathbf F$ and $\mathbf n^{\rm F}=\mathbf0$.  Eq.~\eqref{eq:stress_tensor} gives
\begin{equation}
\mathbf t = \operatorname{diag}\left(
C_1\lambda_1^2(\lambda_1^2-1)-p,\,
2C_4\lambda_2^2(\lambda_2^2-1)-p,\,
2C_4\lambda_3^2(\lambda_3^2-1)-p
\right).
\end{equation}
The lateral condition $t_{22}-t_{33}=0$ therefore derives
\begin{equation*}
2C_4(\lambda_2^2-\lambda_3^2)
(\lambda_2^2+\lambda_3^2-1)=0.
\end{equation*}
The first factor, together with $\lambda_3=(\lambda_1\lambda_2)^{-1}$, gives the equal-contraction solution $\lambda_2=\lambda_3=\lambda_1^{-1/2}$. Eliminating $p$ with $t_{33}=0$ and evaluating $P_{11}=(t_{11}-t_{33})/\lambda_1$ then yields
\begin{equation}
\label{ch3:eq:parallel_response}
P_\parallel
=C_1\lambda_1(\lambda_1^2-1)
-2C_4\lambda_1^{-2}(\lambda_1^{-1}-1).
\end{equation}
The semisoft energy and its stress contribution vanish on this branch, so its response is governed solely by ordinary network elasticity.

The second factor gives $\lambda_2^2+\lambda_3^2=1$. Using $\lambda_2^2\lambda_3^2=\lambda_1^{-2}$ gives the unequal transverse contractions
\begin{equation}
\label{ch4:eq:parallel_transverse_branches}
\lambda_2^2=\frac{1\pm\sqrt{1-4/\lambda_1^2}}2,\qquad
\lambda_3^2=\frac{1\mp\sqrt{1-4/\lambda_1^2}}2.
\end{equation}
These branches coincide at $\lambda_1=2$. When $\lambda_1>2$, the symmetric of transverse contractions is broken, i.e. $\lambda_2\ne\lambda_3$.  This bifurcation does not require director rotation.

\subsection{Perpendicular tension}
\label{ch3:subsec:perpendicular}

For $\theta_0=\pi/2$, the reorientation response has three stages: a locked branch A, a rotating branch B and an aligned branch C. We first describe the variant rotating toward $\mathbf e_1$, its mirror variant and their compatible combination are then discussed below.

\subsubsection{Locked branch and onset}
\label{ch3:subsubsec:onset}

In stage A, $\mathbf n=\mathbf n_0=\mathbf e_2$ and $\mathbf n^{\rm F}=\mathbf0$. The lateral stress boundary condition $t_{22}=t_{33}=0$ and the axial nominal stress $t_{11}-t_{33}=\sigma$ reduce to
\begin{equation}
\label{ch3:eq:A_response}
\left\{
\begin{aligned}
&C_1\lambda_2^2(\lambda_2^2-1)
-
2C_4\lambda_3^2(\lambda_3^2-1)
=0, \qquad \lambda_3=(\lambda_1\lambda_2)^{-1},
\\
&P_A = \frac{\sigma}{\lambda_1} =
\frac{2C_4}{\lambda_1}
\left[
\lambda_1^2(\lambda_1^2-1)
-
\lambda_3^2(\lambda_3^2-1)
\right], \qquad 1\leq\lambda_1\leq\lambda_c.
\end{aligned}
\right.
\end{equation}

\begin{figure}[htbp]
\centering
\includegraphics[width=0.98\textwidth,height=0.65\textheight,keepaspectratio]{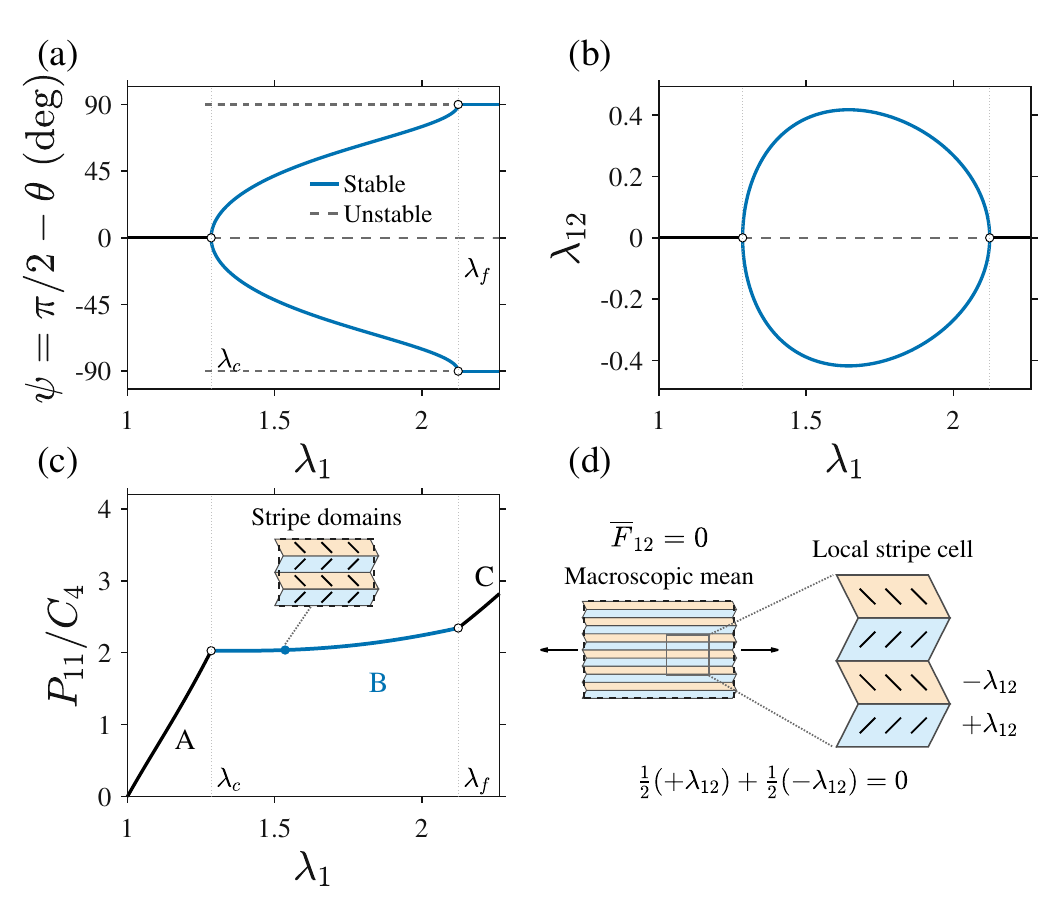}
\caption{
Perpendicular tension ($\theta_0=90^\circ$):
(a) director rotation $\psi=\pi/2-\theta$;
(b) local shear;
(c) nominal stress with a stripe-domain inset;
(d) a compatible equal-fraction laminate with opposite local shear
and zero macroscopic shear.
Solid and dashed curves in (a,b) denote locally stable and unstable
equilibria under homogeneous planar perturbations at fixed
$\lambda_1$; circles mark reorientation onset and completion.
Parameters: $C_1/C_4=10$, $C_5/C_4=3$, $r=3$,
$\alpha_0=0.55$, $\alpha_\infty=0.385$ and $m=0.65$.
}
\label{fig:tensile_bifurcation_symmetry}
\end{figure}
To determine the onset of director rotation, we can denote $\theta=\pi/2-\psi$ by an infinitesimal rotation angle $|\psi|\ll1$. 
The spontaneous stretch tensor can be expanded as
\begin{equation}
\label{eq:expand_spontaneous_stretch}
\begin{aligned}
&\mathbf V^{\rm s} = r^{-1/6}\begin{bmatrix}
1&(\sqrt{r}-1)\psi&0\\
(\sqrt{r}-1)\psi&\sqrt{r}&0\\
0&0&1
\end{bmatrix}+O(\psi^2), \\
&(\mathbf V^{\rm s})^{-1} = r^{1/6} \begin{bmatrix}
1&(\frac{1}{\sqrt{r}}-1)\psi&0\\
(\frac{1}{\sqrt{r}}-1)\psi&\frac{1}{\sqrt{r}}&0\\
0&0&1
\end{bmatrix}+O(\psi^2)
\end{aligned}
\end{equation}
In locking stage, $\mathbf F^{\rm e}= \mathbf F$, the elastic Cauchy stress $\mathbf t^{\rm e} = \mathbf F (\mathbb C^{\rm TI}:\mathbf E)\mathbf F^{\rm T}$ is independent of $\psi$, where $\mathbf E = (\mathbf F^{\rm T}\mathbf F-\mathbf I)/2$ is the Green-Lagrange strain of total deformation. Thus, the shear component of soft elastic Cauchy stress in Eq.~\ref{eq:soft_stress} can be expanded with $\psi$ as follows:
\begin{equation}
\label{eq:expanded_shear_stress_12}
t_{12}^{\rm soft}=\sqrt{r} t_{12} ^{\rm e}-(\sqrt{r}-1)(t^{\rm e}_{11}-t^{\rm e}_{22})\psi+O(\psi^3),
\end{equation}
\begin{equation}
\label{eq:expanded_shear_stress_21}
t_{21}^{\rm soft}=\frac{1}{\sqrt{r}} t_{21} ^{\rm e}-(\frac{1}{\sqrt{r}}-1)(t^{\rm e}_{11}-t^{\rm e}_{22})\psi+O(\psi^3).
\end{equation}
Combining Eq. (\ref{eq:expanded_shear_stress_12}) and Eq. (\ref{eq:expanded_shear_stress_21}), we can remove the shear components of $\mathbf t^{\rm e}$ and derive the following formula
\begin{equation}
\label{eq:critical_stress_conditon_1}
t_{21}^{\rm soft}=\frac{t_{12}^{\rm soft}}{r}-\frac{r-1}{r}(t^{\rm e}_{11}-t^{\rm e}_{22})\psi+O(\psi^3).
\end{equation}
Here, the symmetry of $\mathbf t^{\mathrm e}$ is considered, i. e. $t^{\rm e}_{12}=t^{\rm e}_{21}$. Noticing that $t^{\rm e}_{11}-t^{\rm e}_{22}=t_{11}-t_{22}=\sigma=\lambda_1 P_{A}$, and combining the expanded semisoft Cauchy stress
\begin{equation}
\label{eq:semisoft_shear_components}
t^{\rm semi}_{12} = O(\psi^3), \qquad t^{\rm semi}_{21}=2D_{\rm ss}\lambda_1^2\psi+O(\psi^3),
\end{equation}
we derive the critical condition of total Cauchy stress
\begin{equation}
\label{eq:critical_stress_conditon_2}
t_{21}=\frac{t_{12}}{r}+ (2D_{\rm ss}\lambda_1^2-\frac{r-1}{r}\lambda_1 P_A)\psi+O(\psi^3).
\end{equation}
Substituting the shear boundary condition $t_{12}=0$ and director balance $t_{12}-t_{21}=0$, we obtain the critical nominal stress
\begin{equation}
\label{ch3:eq:onset}
P_A |_{\lambda_1 = \lambda_c}
=
\frac{2r{D}_{\rm ss}}{r-1}
\lambda_c,
\end{equation}
after which the nontrivial rotational mode appears. Combining Eq. (\ref{ch3:eq:onset}) with Eqs. (\ref{ch3:eq:A_response}), we can solve the value of critical stretch $\lambda_c$ and critical stress $P_A |_{\lambda_1 = \lambda_c}$.

\subsubsection{Rotating branch}
\label{ch3:subsubsec:rotation}
In stage B, $0<\theta<\pi/2$. The lateral traction condition $t_{22}-t_{33}=0$, shear traction condition $t_{21}=0$, and angular-momentum balance $t_{12}-t_{21}=0$ determine $\lambda_2$, $\lambda_{12}$, and $\theta$, for each prescribed $\lambda_1\in[\lambda_c,\lambda_f]$.  An equivalent stress expression follows by resolving the stress along $\mathbf n = \cos\theta\,\mathbf e_1 +\sin\theta\,\mathbf e_2$ and $\mathbf n_\perp=-\sin\theta\,\mathbf e_1 +\cos\theta\,\mathbf e_2$.

Using the symmetric of $\mathbf t^{\rm e}$, we have $\mathbf n \cdot\mathbf t^{\rm e}\mathbf n_\perp=\mathbf n_\perp \cdot\mathbf t^{\rm e}\mathbf n$, i.e. $\mathbf n \cdot \mathbf V^{\rm s}\mathbf t^{\rm soft}(\mathbf V^{\rm s})^{-1} \mathbf n_\perp = 
\mathbf n_\perp \cdot \mathbf V^{\rm s}\mathbf t^{\rm soft}(\mathbf V^{\rm s})^{-1} \mathbf n$, by which we derive
\begin{equation}
\label{eq:soft_shear_relation}
\begin{aligned}
r\,\mathbf n\cdot(\mathbf t-\mathbf t^{\rm semi})\mathbf n_\perp
=\mathbf n_\perp\cdot(\mathbf t-\mathbf t^{\rm semi})\mathbf n.
\end{aligned}
\end{equation}
The hydrostatic pressure $p$ drops out because $\mathbf n\cdot\mathbf n_\perp=0$. Substituting $\mathbf t=\lambda_1 P_B\,\mathbf e_1\otimes\mathbf e_1$ and $\mathbf n^{\rm F}=\lambda_1\cos\theta\,\mathbf e_1$ into Eq. (\ref{eq:soft_shear_relation}) yields
\begin{equation}
\begin{aligned}
0={}&\left[(r-1)\lambda_1 P_B
-2rD_{\rm ss}\lambda_1^2\right]\sin\theta\cos\theta\\
&+2(r-1)D_{\rm ss}'\lambda_1^2\cos^2\theta\,
\mathbf n\cdot\mathbf F\mathbf F^{\rm T}\mathbf n_\perp.
\end{aligned}
\end{equation}
Dividing by $(r-1)\lambda_1\sin\theta\cos\theta$ gives
\begin{equation}
\label{ch3:eq:PB}
\begin{aligned}
P_B={}&\frac{2rD_{\rm ss}}{r-1}\lambda_1-2D_{\rm ss}'\lambda_1\cot\theta\, \mathbf n\cdot\mathbf F\mathbf F^{\rm T}\mathbf n_\perp .
\end{aligned}
\end{equation}
The first term in Eq.~\eqref{ch3:eq:PB} depends on the current semisoft modulus, which is consistent with the classical semisoft modification \cite{li2025extended}, while the second term accounts for its dependence on $I_1$.

\subsubsection{Aligned branch and completion}
\label{ch3:subsubsec:completion}

For the continuous branch shown in
Fig.~\ref{fig:tensile_bifurcation_symmetry},
$\lambda_f$ is the stretch at which the rotating solution
reaches $\theta=0$ and $\lambda_{12}=0$.
Stage C then has $\theta=0$ (equivalently $\theta=\pi$)
and $\lambda_{12}=0$. $\mathbf F^{\rm e}
=\operatorname{diag}(\lambda_1/r^{1/2},r^{1/2}\lambda_2,\lambda_3)$ and $\mathbf n^{\rm F}=\lambda_1\mathbf e_1$. The lateral traction condition $t_{22}-t_{33}=0$ and nominal stress $P_C=(t_{11}-t_{33})/\lambda_1$ give
\begin{equation}
\label{ch3:eq:C_response}
\left\{
\begin{aligned}
&C_1r\lambda_2^2(r\lambda_2^2-1)
-2C_4\lambda_3^2(\lambda_3^2-1)
+2D_{\rm ss}'\lambda_1^2(\lambda_2^2-\lambda_3^2)=0,\\
&P_C=\frac{2C_4}{\lambda_1}
\left[\frac{\lambda_1^2}{r}\left(\frac{\lambda_1^2}{r}-1\right)
-\lambda_3^2(\lambda_3^2-1)\right]+2D_{\rm ss}\lambda_1
+2D_{\rm ss}'\lambda_1(\lambda_1^2-\lambda_3^2).
\end{aligned}
\right.
\end{equation}
The first equation determines $\lambda_2$ at a prescribed $\lambda_1$, with $\lambda_3=(\lambda_1\lambda_2)^{-1}$, the second equation then gives the uniaxial stress.

\subsubsection{Mirror variants and compatible stripe domains}
\label{ch4:subsec:stripe_domains}

The perpendicular problem is invariant under transition $(\lambda_{12},\psi)\mapsto(-\lambda_{12},-\psi)$, where $\psi=\pi/2-\theta$. Accordingly, the locked branch bifurcates into two equal-energy director and shear variants, illustrated in Fig.~\ref{fig:tensile_bifurcation_symmetry}(a,b). Local stability in Fig.~\ref{fig:tensile_bifurcation_symmetry}(a,b) is evaluated from the second variation of the homogeneous energy with respect to $\lambda_2$, $\lambda_{12}$ and $\theta$ at fixed $\lambda_1$, with incompressibility enforced. At a common $\lambda_1$ in stage B, the variants can be written as
\begin{equation}
\label{ch4:eq:mirror_variants}
\begin{aligned}
\mathbf F^\pm
&=\operatorname{diag}(\lambda_1,\lambda_2,\lambda_3)
\pm\lambda_{12}\,\mathbf e_1\otimes\mathbf e_2,\\
\mathbf n^\pm
&=\pm\cos\theta\,\mathbf e_1+\sin\theta\,\mathbf e_2.
\end{aligned}
\end{equation}
They share $\lambda_2,\lambda_3$ and $P_B$, but have opposite local shear and opposite rotation senses.

These states also satisfy the geometric compatibility condition for a planar interface with reference normal
\begin{equation}
\label{ch4:eq:stripe_compatibility}
\mathbf F^+-\mathbf F^-
=2\lambda_{12}\,\mathbf e_1\otimes\mathbf N,\qquad
(\mathbf P^+-\mathbf P^-)\mathbf N=\mathbf0.
\end{equation}
Here, $\mathbf N=\mathbf e_2$. The second relation follows because the uniaxial traction conditions give $\mathbf P^\pm=P_B\,\mathbf e_1\otimes\mathbf e_1$. Thus alternating layers can be joined continuously while maintaining traction continuity. For equal volume fractions,
\begin{equation}
\label{ch4:eq:stripe_average}
\overline{\mathbf F}
=\tfrac12(\mathbf F^++\mathbf F^-)
=\operatorname{diag}(\lambda_1,\lambda_2,\lambda_3),
\qquad
\overline P_{11}=P_B.
\end{equation}
The local shear therefore cancels macroscopically without changing the tensile stress, whose three stages are shown in Fig.~\ref{fig:tensile_bifurcation_symmetry}(c). The inset in panel (c) and the local enlargement in panel (d) illustrate the alternating mirror variants, the dashed rectangular outline in panel (d) denotes their shear-free macroscopic mean. This construction provides an idealized way to accommodate macroscopic shear-free extension through local director rotation and shear.

\subsection{Oblique tension}
\label{ch3:subsec:oblique}

For $0<\theta_0<\pi/2$, the semisoft distortion is
\begin{equation}
\label{ch4:eq:oblique_distortion}
|\mathbf n^{\rm F}|^2
=\left[
(-\lambda_1\sin\theta_0+\lambda_{12}\cos\theta_0)\cos\theta
+\lambda_2\cos\theta_0\sin\theta
\right]^2.
\end{equation}
Director rotation, shear and contraction are coupled from the beginning of loading, and the last three equations in Eq.~\eqref{ch3:eq:governing_equations} are solved simultaneously. The initial tensile slope is $E(\theta_0)$ in Eq.~\eqref{ch3:eq:Etheta}.

The angular bias also removes the equivalence of the two perpendicular rotation senses. Set $\eta=\theta_0-\pi/2$ and $\psi=\pi/2-\theta$. The energy is invariant under the combined transformation $(\lambda_{12},\psi,\eta)\mapsto (-\lambda_{12},-\psi,-\eta)$. At a fixed $\eta\ne0$, reflection of the current rotation therefore no longer preserves the reference orientation. The oblique responses below illustrate the resulting bias: director rotation begins from the initial loading stage and continues toward the tensile axis.

\begin{figure}[htbp]
\centering
\includegraphics[
width=0.98\textwidth,
height=0.67\textheight,
keepaspectratio
]{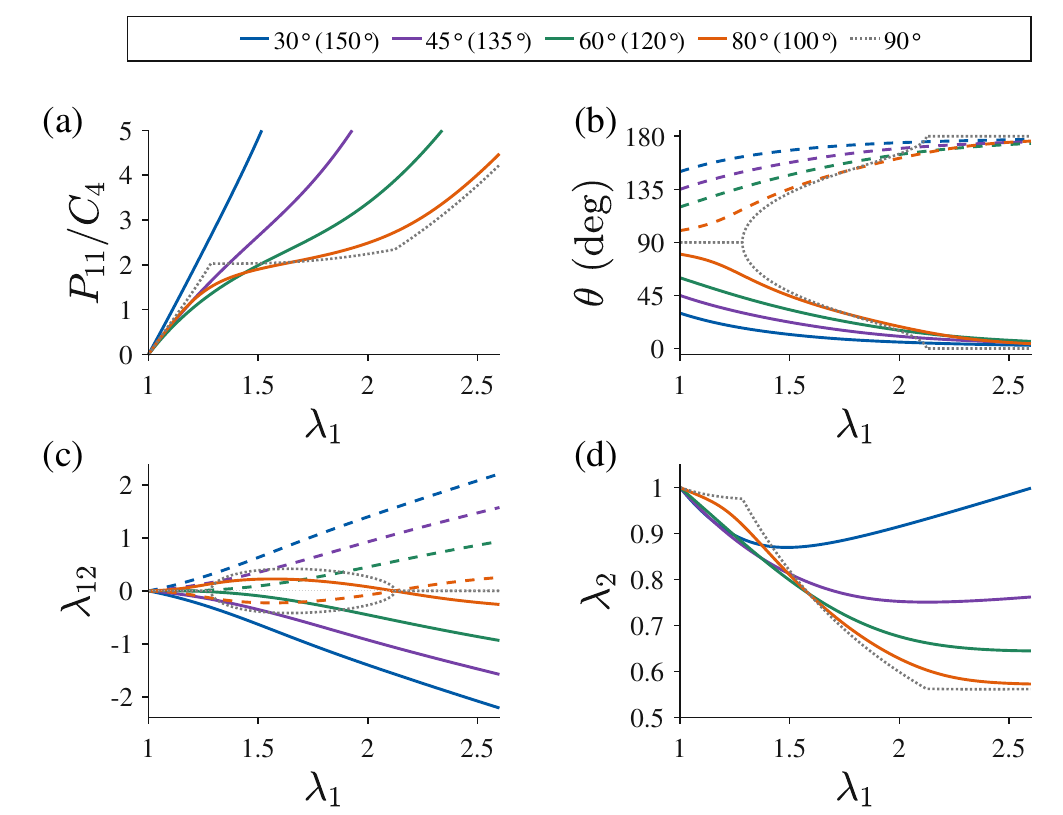}
\caption{
Orientation effects under free-shear uniaxial tension,
using the material parameters of
Fig.~\ref{fig:tensile_bifurcation_symmetry}.
(a) Normalized nominal stress $P_{11}/C_4$.
(b) Director angle $\theta$.
(c) Local shear $\lambda_{12}$.
(d) Lateral stretch $\lambda_2$.
Solid curves correspond to
$\theta_0=30^\circ$, $45^\circ$, $60^\circ$ and $80^\circ$.
In (b,c), dashed curves of the same colors show the
complementary initial angles
$150^\circ$, $135^\circ$, $120^\circ$ and $100^\circ$,
respectively.
Gray dotted curves show perpendicular tension, including
both mirror variants in (b,c).
The reflected responses have coincident stress and lateral
stretch in (a,d).
}
\label{fig:oblique_tension_fields}
\end{figure}

Figure~\ref{fig:oblique_tension_fields} compares
$\theta_0=30^\circ$, $45^\circ$, $60^\circ$ and $80^\circ$
using the same material parameters as
Fig.~\ref{fig:tensile_bifurcation_symmetry}.
The perpendicular response is included as a reference.
Panels (a) and (b) show that changing $\theta_0$ modifies both
the tensile response and the interval over which director
rotation accommodates extension.
The $30^\circ$ response develops the strongest late hardening,
whereas the $80^\circ$ response retains a broader interval of
low tensile slope and rapid rotation.
The initial stiffness depends nonmonotonically on $\theta_0$,
as predicted by Eq.~\eqref{ch3:eq:Etheta}.
For all four oblique cases, the director angle decreases
continuously over the plotted interval.

The shear curves distinguish changes that the stress alone
does not reveal. At $30^\circ$ and $45^\circ$,
$\lambda_{12}<0$ over the plotted loading interval.
At $80^\circ$, a larger positive excursion is followed by a
later reversal.
These sign changes occur while $\theta$ continues to decrease;
they do not indicate a reversal of director rotation.
The initial shear directions reflect the anisotropic linear
compliances in Appendix~\ref{app:linear_compatibility}.
Finally, panel (d) shows that $\lambda_2$ need not decrease
monotonically: the $30^\circ$ and $45^\circ$ paths contract
and then recover, whereas the $80^\circ$ path sustains
stronger contraction over this interval.

The complementary initial orientation $\pi-\theta_0$ gives
the reflected response
$(\theta,\lambda_{12})\mapsto(\pi-\theta,-\lambda_{12})$
at the same $\lambda_1$, while $P_{11}$ and $\lambda_2$
are unchanged.
The solid--dashed pairs in
Fig.~\ref{fig:oblique_tension_fields}(b,c)
therefore refer to complementary initial orientations.
At $\theta_0=90^\circ$, this reflection connects the two
mirror variants of the same perpendicular-loading problem.

\section{Material control and comparison with tensile experiments}
\label{sec:material_control}

The preceding analysis separates initial network stiffness from
finite-deformation resistance to director reorientation. We now
illustrate the resulting material controls and assess the tensile
response across loading orientations.

\subsection{Network anisotropy and finite-strain response}
\label{ch5:subsec:network_control}

\begin{figure}[htbp]
\centering
\includegraphics[
width=0.98\textwidth,
height=0.60\textheight,
keepaspectratio
]{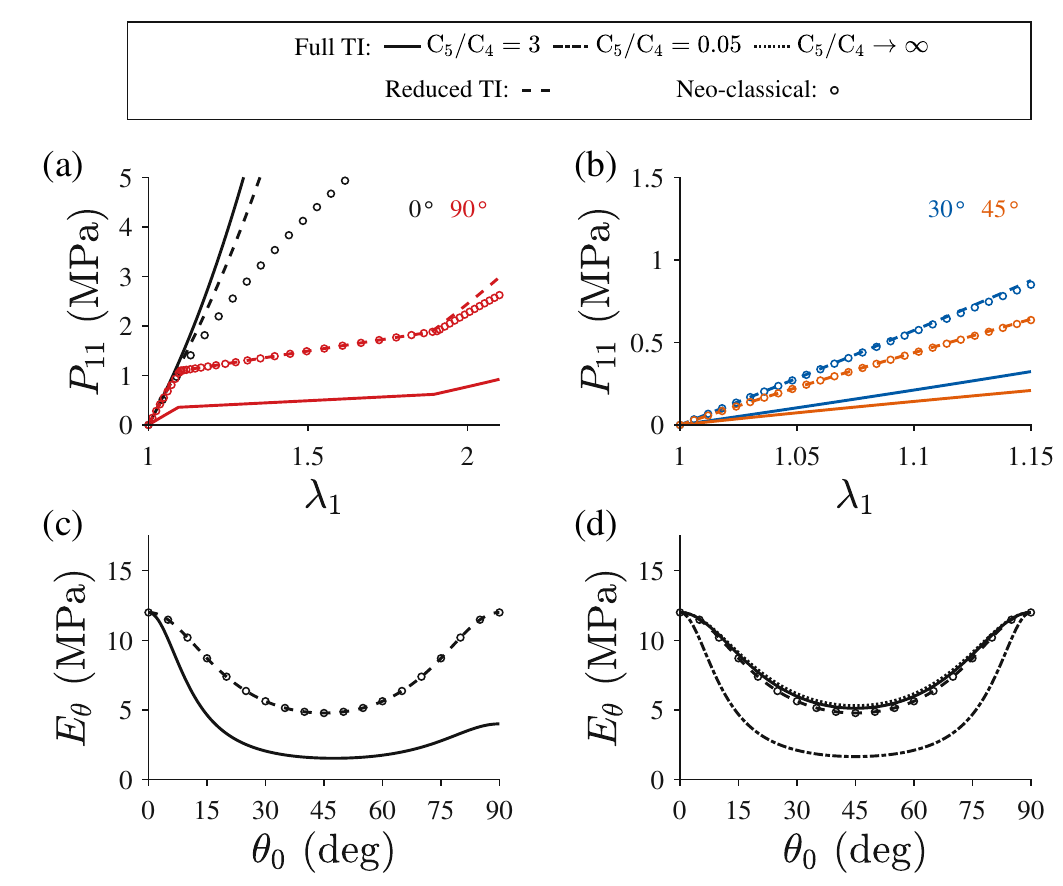}
\caption{
Network contributions to directional stiffness and finite-strain tension, with a constant semisoft modulus for each parameter set. Parameters are listed in Table~\ref{tab:soft_elasticity_comparison_parameters}. (a) Principal tensile responses for $\theta_0=0^\circ$ (black) and $90^\circ$ (red). (b) Oblique responses for $\theta_0=30^\circ$ (blue) and $45^\circ$ (orange). Solid curves in (a--c) show the full TI model. (c) Initial angular modulus for the anisotropic TI example and the matched reduced TI and neo-classical references. (d) Full TI parameter variations at fixed $C_1=2C_4$: solid, dash-dotted and dotted curves use $C_5/C_4=3$, $0.05$ and $+\infty$, respectively. }
\label{fig:soft_elasticity_comparison}
\end{figure}

The principal stiffness contrast and the compliance at oblique angles impose distinct requirements on the elastic description. We first examine these requirements with $D_{\rm ss}$ held constant for each parameter set, before considering its deformation dependence. The fixed-$r$ neo-classical model gives $E_\parallel=E_\perp=3\mu$, whereas Eq.~\eqref{eq:SI_apparent_moduli} allows the transversely isotropic (TI) network to represent unequal principal moduli through $C_1/C_4$. In Fig.~\ref{fig:soft_elasticity_comparison}(c), the full TI example retains $E_\parallel\simeq12$ MPa while reducing $E_\perp$ to approximately $4$ MPa. The principal modulus ratio therefore supplies a direct constraint on the network anisotropy.

The angular stiffness additionally depends on the director-relaxed shear modulus $G_{\rm d}$ through Eq.~\eqref{ch3:eq:Etheta}. At fixed $C_1,C_4,r$ and $D_{\rm ss}$, varying $C_5$ changes this shear modulus while leaving both principal Young's moduli unchanged. Panel (d) isolates this contribution: increasing $C_5$ raises the intermediate-angle stiffness, with the principal values remaining fixed. The effect saturates as $C_5\to\infty$, consistently with the finite bound in Eq.~\eqref{ch3:eq:shear_modulus_bound}.
Thus, the principal stiffness contrast and the oblique compliance are governed by distinct combinations of the network coefficients.

Matching the initial response does not, however, determine the finite-strain tensile behavior. To examine this distinction, the reduced TI and neo-classical references use the same $r$ and $D_{\rm ss}$, with $C_1=\mu$ and $C_4=C_5=\mu/2$ for the reduced TI case. These choices match their complete quadratic response. Their network energies also retain the same ideal soft-mode family, but assign different energies to elastic distortions away from that family. The differences between their tensile curves in panels (a,b) therefore identify the contribution of the nonlinear network energy, including its effect on the coupled oblique response. The full TI curves show the response of the anisotropic parameter set in Table~\ref{tab:soft_elasticity_comparison_parameters}, whose constant semisoft coefficient also differs from that of the matched references. The parameter choices and matching conditions are detailed in Appendix~\ref{appD:model_matching}.

These comparisons show that the initial moduli and ideal soft modes do not uniquely determine finite-strain tension. For prescribed network parameters and $r$, a further constitutive requirement concerns the relation between reorientation onset and subsequent stress growth. The role of the semisoft law in controlling this relation is examined next.

\subsection{Semisoft control of reorientation onset and stress growth}
\label{ch3:subsubsec:semisoft_parameter_roles}

\begin{figure}[htbp]
\centering
\includegraphics[
width=0.98\textwidth,
height=0.58\textheight,
keepaspectratio
]{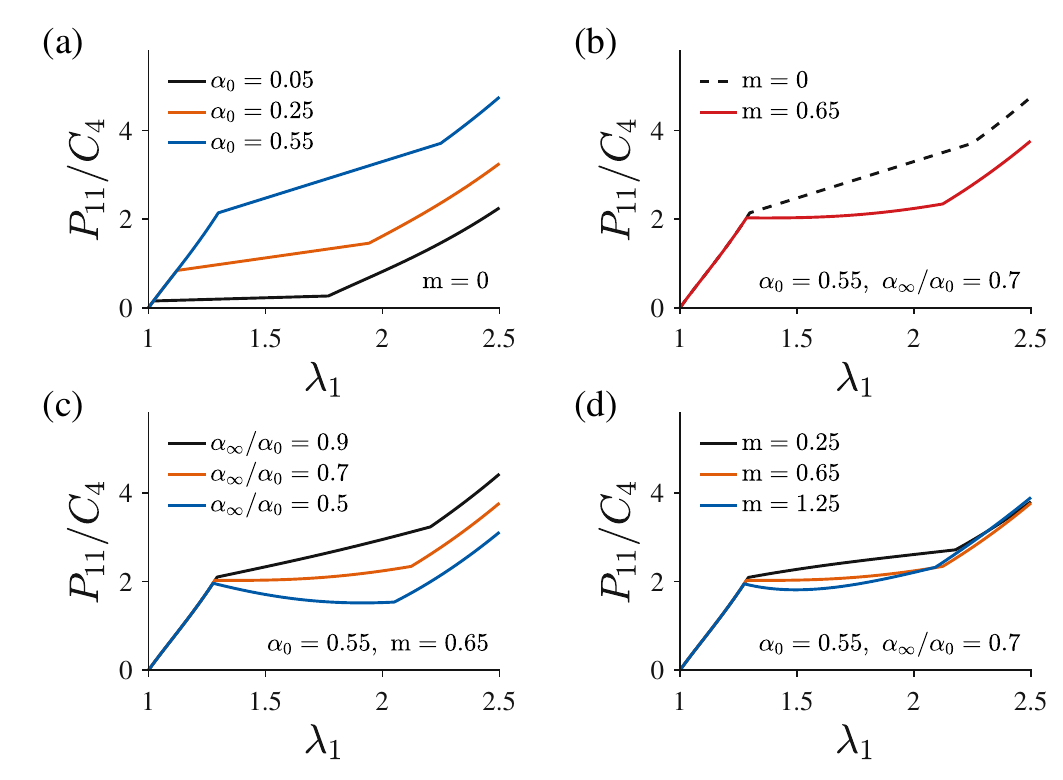}
\caption{
Semisoft control of perpendicular homogeneous tension at
$C_1=10C_4$, $C_5=3C_4$ and $r=3$.
All panels show normalized nominal stress $P_{11}/C_4$
against $\lambda_1$.
(a) Constant coefficient, $m=0$, with
$\alpha_0=0.05$, $0.25$ and $0.55$.
(b) Constant and deformation-dependent coefficients at
$\alpha_0=0.55$; the red curve uses
$\alpha_\infty/\alpha_0=0.7$ and $m=0.65$.
(c) Residual-coefficient variation at
$\alpha_0=0.55$ and $m=0.65$, with
$\alpha_\infty/\alpha_0=0.9$, $0.7$ and $0.5$.
(d) Decay-scale variation at
$\alpha_0=0.55$ and $\alpha_\infty/\alpha_0=0.7$,
with $m=0.25$, $0.65$ and $1.25$.
}
\label{fig:semisoft_decay_comparison}
\end{figure}

With the network parameters and $r$ specified, perpendicular
tension provides an analytical test of whether a finite
reorientation onset can be followed by weak stress growth.
For a constant semisoft coefficient, eliminating $D_{\rm ss}$
between the onset condition~\eqref{ch3:eq:onset} and the
rotating-stress relation~\eqref{ch3:eq:PB} gives
\begin{equation}
\label{ch5:eq:constant_modulus_coupling}
P_B(\lambda_1)
=
\frac{P_A(\lambda_c)}{\lambda_c}\lambda_1.
\end{equation}
Thus, once the locked response and onset stretch are prescribed,
the rotating-stage slope is fixed.
Figure~\ref{fig:semisoft_decay_comparison}(a) illustrates this
restriction: increasing the constant semisoft resistance delays
rotation but also increases the subsequent stress growth.

The deformation dependence in Eq.~\eqref{eq:semisoft_law}
provides additional control of this relation.
We use $\alpha_0=D_{\rm ss}^{0}/C_4$ and
$\alpha_\infty=D_{\rm ss}^{\infty}/C_4$.
At fixed network parameters, $r$ and $\alpha_0$, varying
$\alpha_\infty$ and $m$ leaves the initial quadratic energy
unchanged.
Moreover, $\mathbf n^{\rm F}=\mathbf0$ on the perpendicular
locked branch, so these variations preserve its stress
relation $P_A(\lambda_1)$.
Panel (b) shows that these common initial properties can
coexist with a substantially flatter rotating response when
the semisoft coefficient decreases with deformation.

This change follows from both the current coefficient
$D_{\rm ss}$ and its derivative with respect to $I_1$.
In Eq.~\eqref{ch3:eq:PB}, the first term reflects the current
semisoft resistance, whereas the second arises from the
deformation dependence of that resistance when the energy is
differentiated.
Both contributions are therefore required to evaluate the
rotating stress.
Their combined effect allows a finite pre-rotation extension
to be followed by a low-slope regime, beyond the linear
relation imposed by a constant coefficient.

Panels (c,d) distinguish the magnitude of the available
reduction from the deformation scale over which it develops.
At fixed $\alpha_0$, lowering $\alpha_\infty$ increases this
reduction and can produce a locally decreasing portion of the
homogeneous rotating branch.
The parameter $m$ controls how rapidly the coefficient
approaches its residual value as $I_1$ increases.
The curve crossings in panel (d) show that changing this
scale affects different portions of the tensile response
differently; increasing $m$ does not lower the stress
throughout the loading path.

The onset itself depends on the current coefficient at
$\lambda_1=\lambda_c$ through
Eq.~\eqref{ch3:eq:onset}.
Consequently, changing $\alpha_\infty$ or $m$ can shift
$\lambda_c$ even when the initial moduli are unchanged.
The three semisoft parameters therefore jointly control
the onset and subsequent stress development through a
single coefficient law.
The perpendicular onset condition is independent of $C_5$,
whose role in the oblique shear compliance was examined above.
The experimental comparison below assesses whether these
constitutive controls can describe the tensile response
across loading orientations using a common parameter set.

\subsection{Comparison with tensile test experiments}
\label{ch3:subsec:experimental_comparison}

\begin{figure}[p]
\centering
\includegraphics[
width=0.98\textwidth,
height=0.72\textheight,
keepaspectratio
]{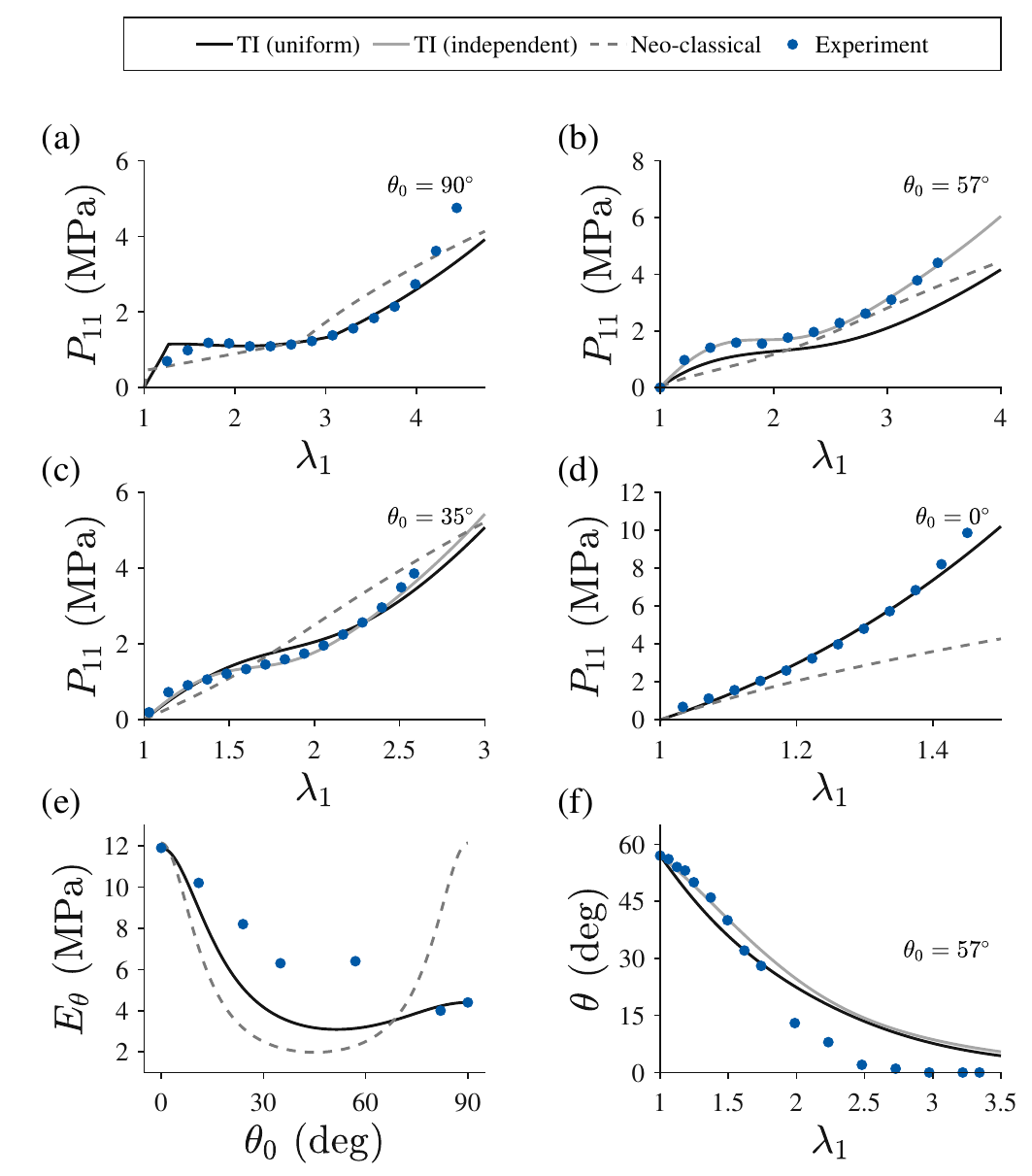}
\caption{
Comparison with \citet{okamoto2021effect}.
(a)--(d) Nominal stress at
$\theta_0=90^\circ$, $57^\circ$, $35^\circ$ and $0^\circ$.
(e) Initial angular modulus $E_\theta$.
(f) Director evolution at $\theta_0=57^\circ$.
Black and light-gray solid curves denote the common and
angle-specific TI sets; gray dashed curves denote the
neo-classical reference, and blue symbols show measurements.
The $57^\circ$ and $82^\circ$ stiffnesses are estimated
from Figs.~7(b) and 3(c) of that study; director data are
digitized from Fig.~7(b).
Parameters are listed in
Table~\ref{appD:tab:experimental_parameters}.
}
\label{fig:stress-curves-initial-angles}
\end{figure}

The experimental requirement is a common description of
directional stiffness and finite-strain reorientation.
Figure~\ref{fig:stress-curves-initial-angles} examines this
requirement using the measurements of
\citet{okamoto2021effect}.
The common TI parameter set, labeled \emph{uniform}, uses
$E_\parallel=11.9$ MPa and $E_\perp=4.4$ MPa to determine
$C_1$ and $C_4$ through
Eq.~\eqref{eq:SI_apparent_moduli}.
With $C_5=C_4$ imposed, the remaining parameters were
selected by manual adjustment against the tensile stress
curves and then held fixed across loading orientations.
The parameter sets and calibration procedure are reported
in Table~\ref{appD:tab:experimental_parameters} and
Appendix~\ref{app:parameter_identification}.

The principal stiffness contrast is therefore incorporated
as a calibration constraint.
The fixed-$r$ neo-classical reference instead gives
$E_\parallel=E_\perp=3\mu=12.14$ MPa and cannot
represent both principal values in panel (e).
The parallel stress curve in panel (d) provides a further
assessment of the finite-strain network energy.
Since the semisoft stress vanishes on this branch, its
nonlinear hardening directly reflects the selected network
response.

With the same TI parameters, perpendicular tension exhibits
a finite onset followed by a broad interval of weak stress
growth, panel (a).
This is the combination targeted by the semisoft
deformation dependence examined in
Section~\ref{ch3:subsubsec:semisoft_parameter_roles}.
The common set represents this broad soft response, although
both the TI and neo-classical calculations underestimate
the stronger post-plateau hardening at the largest stretches.

Oblique loading tests the simultaneous description of
angular stiffness and the coupled finite-strain response.
The common set captures the main curvature changes at
$35^\circ$, but underestimates much of the intermediate
and large-deformation stress at $57^\circ$.
Several intermediate-angle moduli in panel (e) also exceed
the calculated values.
The separate $35^\circ$ and $57^\circ$ calibrations,
labeled \emph{independent}, improve the stress comparison
while retaining $C_1$ and $C_4$, and hence both principal
moduli.
They adjust the shear, anisotropy and semisoft parameters.
These curves illustrate the improvement obtainable by
separate calibration; the common set remains the basis
for assessing agreement across orientations.

Panel (f) examines the associated director evolution using
measurements excluded from parameter selection.
Both theoretical calculations reproduce a continuous decrease in $\theta$, but retain larger angles than measured for
$\lambda_1\gtrsim2$.
The experimental director is nearly aligned by
$\lambda_1\simeq2.7$, while both calculations retain a
residual inclination.
For the reported parameter sets, improving the
$57^\circ$ stress fit therefore leaves a substantial
late-stage discrepancy in the rotation path.
The director measurements supply an additional constraint
that is not resolved by stress agreement alone.

Joint calibration against stress and director measurements is needed to determine how far this discrepancy can be reduced within the present energy. This assessment must also account for the parameter
combinations constrained by each observable: the initial angular modulus determines $G_{\rm d}$,
which combines $C_5$ and $D_{\rm ss}^{0}$, rather than
identifying them separately.
Measurements of lateral contraction and shear would
further constrain the coupled deformation underlying
the tensile response.

\section{Conclusions}
\label{sec:conclusions}

We have developed an isothermal elastic model for the
orientation-dependent response of monodomain main-chain
liquid crystal elastomers by combining transversely
isotropic network elasticity with deformation-dependent
semisoft resistance.
The deformation dependence is motivated by a reduced
statistical closure that assumes contraction of effective
local anisotropy fluctuations at fixed mean chain anisotropy.
The formulation retains the ideal soft modes while allowing
the resistance to director reorientation to change with
finite deformation.

The analytical results clarify the connection between
reorientation onset and subsequent stress growth.
For a given network response, a constant semisoft resistance
links the onset to a linearly increasing nominal stress
during perpendicular rotation.
Allowing this resistance to vary with deformation relaxes
that restriction: substantial extension before rotation
can coexist with a broad, low-slope rotating regime while
preserving the prescribed initial elastic response.
The rotating stress depends on both the current resistance
and how it changes with deformation, making this variation
an essential part of the constitutive description.

The network comparisons also show that identical initial
elasticity and ideal soft modes are insufficient to
determine finite-strain tension.
Different nonlinear energies assigned to distortions away
from the soft modes produce different tensile responses.
This distinction also affects oblique loading, where
director rotation, shear and lateral contraction are
coupled.
Using the principal moduli as calibration inputs, a common
parameter set describes the nonlinear parallel response
and the broad perpendicular soft regime.
Quantitative differences remain in oblique stresses,
angular stiffnesses and large-stretch hardening.
Director measurements excluded from parameter selection
provide an additional test: the reported stress-based
calibrations overestimate the remaining director inclination
at large extension, even when the oblique stress fit is
improved.
Agreement in tensile stress therefore provides only part
of the experimental assessment of the coupled response.

These results link directional stiffness, the onset of director rotation, and plateau development to network anisotropy and evolving semisoft resistance. Future comparisons with combined measurements of stress, director orientation, and lateral deformation will help refine this homogeneous elastic description and establish its range of applicability. Extending the framework to include dissipation would also enable investigation of anisotropic viscoelastic behaviour in LCEs \cite{yang2026thermomechanical}. Together, these findings provide a foundation for understanding the anisotropic response of main-chain LCEs, from initial elasticity to director-mediated deformation at large strains, and may guide the design of soft actuators, compliant structures, and related devices that exploit directional mechanical properties \cite{dai2025unusual}.

\section*{Acknowledgments}
We acknowledge the
financial support from the National Natural Science Foundation of China (Grant No. 12472061).

\appendix
\setcounter{equation}{0} 
\renewcommand{\theequation}{A.{\arabic{equation}}}
\section{Details of the continuum formulation}
\label{app:continuum_details}

This appendix derives the equations used in
Section~\ref{sec:constitutive_model}. Following the power-based
formulations of structured continua
\citep{gurtin1992formulation,green1995unified}, the deformation
and director are assigned independent rates and conjugate forces.
We specialize to an isothermal, quasistatic, nondissipative solid
without external body forces or body couples. Director-gradient
terms are retained temporarily; setting them to zero recovers the
local theory used in the main text.

\subsection{Energy balance and constraint power}
\label{app:continuum_energy_balance}

Throughout this appendix, $\mathbf n$ denotes the current director
evaluated at $\mathbf y=\varphi_t(\mathbf x)$. Its dot denotes the
material time derivative, so that
$\dot{\mathbf F}=\nabla_{\mathbf x}\dot{\mathbf y}$ and
$\dot{\nabla_{\mathbf x}\mathbf n}
=\nabla_{\mathbf x}\dot{\mathbf n}$.
With the reference director $\mathbf n_0$ fixed, write
\begin{equation}
\label{eq:general_free_energy_2}
W=\hat W\left(
\mathbf F,\mathbf n,\nabla_{\mathbf x}\mathbf n;\mathbf n_0
\right),
\qquad
J:=\det\mathbf F=1,
\qquad
M:=\mathbf n\cdot\mathbf n=1.
\end{equation}
For an arbitrary material volume $V_0\subset\mathcal B_0$, the
chain rule gives
\begin{equation}
\label{eq:free_energy_rate}
\dot{\mathcal F}
=\int_{V_0}\left[
\frac{\partial W}{\partial\mathbf F}:\dot{\mathbf F}
+\frac{\partial W}{\partial\mathbf n}\cdot\dot{\mathbf n}
+\frac{\partial W}{\partial\nabla_{\mathbf x}\mathbf n}:
\nabla_{\mathbf x}\dot{\mathbf n}
\right]\,{\rm d}V.
\end{equation}

The multipliers $p$ and $\gamma$ enforce the constraints through
the augmented term
\begin{equation}
\label{eq:constraint_energy}
W^{\rm con}=-p(J-1)-\frac{\gamma}{2}(M-1).
\end{equation}
Using $\dot J=J\mathbf F^{-\rm T}:\dot{\mathbf F}$ and
$\dot M=2\mathbf n\cdot\dot{\mathbf n}$, its constraint power at
$J=M=1$ is
\begin{equation}
\label{eq:constraint_power}
\dot{\mathcal F}^{\rm con}
=\int_{V_0}\left(
-p\mathbf F^{-\rm T}:\dot{\mathbf F}
-\gamma\mathbf n\cdot\dot{\mathbf n}
\right)\,{\rm d}V.
\end{equation}
This power vanishes for rates satisfying the constraints.
The augmented term serves to identify their reaction forces.

The boundary power includes both ordinary and director tractions,
\begin{equation}
\label{eq:external_power}
\dot{\mathcal F}^{\rm ext}
=\int_{\partial V_0}\left(
\mathbf P_{\mathbf N}\cdot\dot{\mathbf y}
+\boldsymbol{\pi}_{\mathbf N}\cdot\dot{\mathbf n}
\right)\,{\rm d}S,
\end{equation}
where $\mathbf N$ is the outward unit normal in the reference
configuration. The nondissipative energy balance is therefore
\begin{equation}
\label{eq:energy_balance}
\dot{\mathcal F}+\dot{\mathcal F}^{\rm con}
=\dot{\mathcal F}^{\rm ext}.
\end{equation}

\subsection{Virtual power and local field equations}
\label{app:continuum_virtual_power}

At a fixed state, the equilibrium problem is expressed by the
augmented principle of virtual power. Introduce independent
virtual rates
\begin{equation}
\label{eq:rate_variations}
\dot{\mathbf y}^{*}=\dot{\mathbf y}+\zeta\mathbf c,
\qquad
\dot{\mathbf n}^{*}=\dot{\mathbf n}+\zeta\mathbf c^{\rm n},
\qquad
\delta\dot{\mathbf y}=\mathbf c,
\quad
\delta\dot{\mathbf n}=\mathbf c^{\rm n},
\end{equation}
where $\delta$ denotes differentiation with respect to $\zeta$ at
$\zeta=0$, with the state and multipliers held fixed. The fields
$\mathbf c$ and $\mathbf c^{\rm n}$ are arbitrary in the augmented
problem. In particular, the director variation is independent of
the deformation variation; it is not restricted to that generated
by a change of observer.

Substituting the energy and constraint rates directly into the
virtual-power statement gives
\begin{equation}
\label{eq:var_energy_balance}
\begin{aligned}
&\int_{V_0}\left[
\left(\frac{\partial W}{\partial\mathbf F}
-p\mathbf F^{-\rm T}\right):\nabla_{\mathbf x}\mathbf c
+\left(\frac{\partial W}{\partial\mathbf n}
-\gamma\mathbf n\right)\cdot\mathbf c^{\rm n}
+\frac{\partial W}{\partial\nabla_{\mathbf x}\mathbf n}:
\nabla_{\mathbf x}\mathbf c^{\rm n}
\right]\,{\rm d}V
\\
&\hspace{2em}=\int_{\partial V_0}\left(
\mathbf P_{\mathbf N}\cdot\mathbf c
+\boldsymbol{\pi}_{\mathbf N}\cdot\mathbf c^{\rm n}
\right)\,{\rm d}S.
\end{aligned}
\end{equation}
This statement for arbitrary virtual rates supplies the equilibrium
equations. Substituting the actual rates recovers the energy
balance~\eqref{eq:energy_balance}.

Integrating the two gradient terms by parts gives
\begin{equation}
\label{eq:var_int_energy_balance}
\begin{aligned}
0={}&\int_{V_0}
\mathrm{Div}\left(
\frac{\partial W}{\partial\mathbf F}-p\mathbf F^{-\rm T}
\right)\cdot\mathbf c\,{\rm d}V
\\
&+\int_{V_0}\left[
\mathrm{Div}\left(
\frac{\partial W}{\partial\nabla_{\mathbf x}\mathbf n}
\right)-\frac{\partial W}{\partial\mathbf n}
+\gamma\mathbf n\right]\cdot\mathbf c^{\rm n}\,{\rm d}V
\\
&+\int_{\partial V_0}\left[
\mathbf P_{\mathbf N}
-\left(\frac{\partial W}{\partial\mathbf F}
-p\mathbf F^{-\rm T}\right)\cdot\mathbf N
\right]\cdot\mathbf c\,{\rm d}S
\\
&+\int_{\partial V_0}\left[
\boldsymbol{\pi}_{\mathbf N}
-\frac{\partial W}{\partial\nabla_{\mathbf x}\mathbf n}
\cdot\mathbf N\right]\cdot\mathbf c^{\rm n}\,{\rm d}S.
\end{aligned}
\end{equation}
First choosing test fields with compact support gives the bulk
equations. Allowing independent boundary values then gives the
traction relations. Together, these are
\begin{equation}
\label{eq:micro_momentum_balance}
\begin{aligned}
\mathrm{Div}\left(
\frac{\partial W}{\partial\mathbf F}-p\mathbf F^{-\rm T}
\right)&=\mathbf0
&&\text{in }V_0,
\\
\mathrm{Div}\left(
\frac{\partial W}{\partial\nabla_{\mathbf x}\mathbf n}
\right)-\frac{\partial W}{\partial\mathbf n}
+\gamma\mathbf n&=\mathbf0
&&\text{in }V_0,
\\
\left(\frac{\partial W}{\partial\mathbf F}
-p\mathbf F^{-\rm T}\right)\cdot\mathbf N
&=\mathbf P_{\mathbf N}
&&\text{on }\partial V_0,
\\
\frac{\partial W}{\partial\nabla_{\mathbf x}\mathbf n}
\cdot\mathbf N&=\boldsymbol{\pi}_{\mathbf N}
&&\text{on }\partial V_0.
\end{aligned}
\end{equation}

Because $V_0$ is arbitrary, the boundary relations hold for every
cut normal $\mathbf N$. The two tensors whose normal traces give
the ordinary and director tractions are therefore identified as
the first Piola--Kirchhoff stress and the director microstress.
The nongradient director term identifies the intrinsic body force.
Thus the derived field and boundary equations identify
\begin{equation}
\label{appA:eq:general_stresses}
\mathbf P=-p\mathbf F^{-\rm T}
+\frac{\partial W}{\partial\mathbf F},
\qquad
\boldsymbol{\pi}
=\frac{\partial W}{\partial\nabla_{\mathbf x}\mathbf n},
\qquad
\mathbf g=\gamma\mathbf n-\frac{\partial W}{\partial\mathbf n}.
\end{equation}
The undifferentiated director term in
Eq.~\eqref{eq:var_energy_balance} is
$-\mathbf g\cdot\mathbf c^{\rm n}$, which fixes the sign convention
for $\mathbf g$. The bulk equations already obtained now read
$\mathrm{Div}\mathbf P=\mathbf0$ and
$\mathrm{Div}\boldsymbol{\pi}+\mathbf g=\mathbf0$.

For the unit director, admissible boundary rates are perpendicular
to $\mathbf n$, so only the tangential component of the director
traction is prescribed physically. The displayed equality selects
its representative, including the normal constraint reaction;
equivalently, the physical condition is
$\mathbf n\times(\boldsymbol{\pi}_{\mathbf N}
-\boldsymbol{\pi}\cdot\mathbf N)=\mathbf0$.

\subsection{Objectivity and angular momentum balance}
\label{app:continuum_angular_momentum}

A superposed rigid rotation acts on the current configuration,
while $\mathbf x$ and $\mathbf n_0$ remain fixed. Its infinitesimal
form is
\begin{equation}
\label{eq:rigid_rotation}
\begin{aligned}
\delta\mathbf y&=\boldsymbol{\Omega}^{\rm skew}\mathbf y,
&\delta\mathbf n&=\boldsymbol{\Omega}^{\rm skew}\mathbf n,
\\
\delta\mathbf F&=\boldsymbol{\Omega}^{\rm skew}\mathbf F,
&\delta(\nabla_{\mathbf x}\mathbf n)
&=\boldsymbol{\Omega}^{\rm skew}
\nabla_{\mathbf x}\mathbf n,
\end{aligned}
\end{equation}
where $\boldsymbol{\Omega}^{\rm skew}$ is arbitrary. Differentiating
the objective energy and using
$\mathbf t=\mathbf P\mathbf F^{\rm T}$ gives
\begin{equation}
\label{appA:eq:Ericksen_identity}
\begin{aligned}
0&=\left(
\frac{\partial W}{\partial\mathbf F}\mathbf F^{\rm T}
+\frac{\partial W}{\partial\mathbf n}\otimes\mathbf n
+\boldsymbol{\pi}(\nabla_{\mathbf x}\mathbf n)^{\rm T}
\right):\boldsymbol{\Omega}^{\rm skew},
\\
\mathbf0&=\boldsymbol{\epsilon}:\left(
\mathbf t
+\frac{\partial W}{\partial\mathbf n}\otimes\mathbf n
+\boldsymbol{\pi}(\nabla_{\mathbf x}\mathbf n)^{\rm T}
\right).
\end{aligned}
\end{equation}
The second line follows because the first holds for every skew
tensor and the pressure term has zero skew part. We use
$(\boldsymbol{\epsilon}:\mathbf A)_i=\epsilon_{ijk}A_{jk}$.
This identity holds independently of director equilibrium.

The director traction produces the surface moment
$\mathbf n\times\boldsymbol{\pi}_{\mathbf N}$. Accordingly,
define the referential couple stress by
\begin{equation}
\label{appA:eq:couple_stress}
\mathbf\Pi:=\boldsymbol{\epsilon}:
(\mathbf n\otimes\boldsymbol{\pi}),
\qquad
\Pi_{iA}=\epsilon_{ijk}n_j\pi_{kA},
\qquad
\mathbf\Pi\cdot\mathbf N
=\mathbf n\times(\boldsymbol{\pi}\cdot\mathbf N).
\end{equation}
Here the capital index $A$ labels a reference coordinate.
The angular momentum balance for a material part is
\begin{equation}
\label{appA:eq:angular_momentum_balance}
\begin{aligned}
\int_{\partial V_0}\left[
\mathbf y\times(\mathbf P\cdot\mathbf N)
+\mathbf\Pi\cdot\mathbf N\right]\,{\rm d}S&=\mathbf0,
\\
\mathrm{Div}\mathbf\Pi-\boldsymbol{\epsilon}:\mathbf t
&=\mathbf0.
\end{aligned}
\end{equation}
To obtain the local form, use
$\mathrm{Div}(\mathbf y\times\mathbf P)
=\mathbf y\times\mathrm{Div}\mathbf P
-\boldsymbol{\epsilon}:\mathbf t$
and the force balance. The same surface statement follows by
using an infinitesimal rigid rotation as the virtual rate in
Eq.~\eqref{eq:var_energy_balance}, with the stresses identified in
Eq.~\eqref{appA:eq:general_stresses}.

The precise relation to director equilibrium follows from the
product rule:
\begin{equation}
\label{appA:eq:couple_stress_divergence}
\mathrm{Div}\mathbf\Pi
=\mathbf n\times\mathrm{Div}\boldsymbol{\pi}
-\boldsymbol{\epsilon}:
\left[\boldsymbol{\pi}
(\nabla_{\mathbf x}\mathbf n)^{\rm T}\right].
\end{equation}
Combining this expression with
Eq.~\eqref{appA:eq:Ericksen_identity} and the definition of
$\mathbf g$ yields
\begin{equation}
\label{appA:eq:balance_equivalence}
\mathrm{Div}\mathbf\Pi-\boldsymbol{\epsilon}:\mathbf t
=\mathbf n\times
(\mathrm{Div}\boldsymbol{\pi}+\mathbf g).
\end{equation}
Thus director equilibrium implies angular momentum balance.
Conversely, angular momentum balance makes the director-balance
residual parallel to $\mathbf n$. Its parallel component is fixed
by the unit-length multiplier,
\begin{equation}
\label{appA:eq:director_multiplier}
\gamma=\mathbf n\cdot\left(
\frac{\partial W}{\partial\mathbf n}
-\mathrm{Div}\boldsymbol{\pi}\right),
\end{equation}
which then gives the full director balance. The equivalence is
therefore understood with the constraint reaction included.

Finally, neglecting the director-gradient energy gives
$\boldsymbol{\pi}=\mathbf0$ and $\mathbf\Pi=\mathbf0$.
The local field equations reduce to
$\mathrm{Div}\mathbf P=\mathbf0$ and $\mathbf g=\mathbf0$,
with the stress and intrinsic body force in
Eq.~\eqref{appA:eq:general_stresses}. In this limit,
$\gamma=\mathbf n\cdot\partial W/\partial\mathbf n$ and
\begin{equation}
\label{appA:eq:local_balance_equivalence}
\mathbf g=\mathbf0
\quad\Longleftrightarrow\quad
\mathbf n\times\frac{\partial W}{\partial\mathbf n}=\mathbf0
\quad\Longleftrightarrow\quad
\boldsymbol{\epsilon}:\mathbf t=\mathbf0
\quad\Longleftrightarrow\quad
\mathbf t=\mathbf t^{\rm T}.
\end{equation}
These are the director and angular momentum equations used in
the main text.

\section{Statistical closure and linearized constitutive relations}
\label{app:linear_compatibility}
\setcounter{equation}{0} 
\renewcommand{\theequation}{B.{\arabic{equation}}}

This appendix gives the probability-transport calculation underlying Eq.~\eqref{eq:semisoft_law} and derives the director-relaxed moduli in Eqs.~\eqref{eq:SI_apparent_moduli} and \eqref{ch3:eq:Etheta}.

\subsection{Statistical closure of the semisoft coefficient}
\label{appB:statistical_closure}

At fixed mean anisotropy $r=\langle r^{(\nu)}\rangle_\nu$, write
$r^{(\nu)}=r+\delta r^{(\nu)}$, with
$\langle\delta r^{(\nu)}\rangle_\nu=0$. For weak fluctuations,
\begin{equation}
\label{appB:eq:inverse_anisotropy_expansion}
\frac1{r^{(\nu)}}
=
\frac1r-\frac{\delta r^{(\nu)}}{r^2}
+\frac{(\delta r^{(\nu)})^2}{r^3}
+O\!\left(\frac{|\delta r^{(\nu)}|^3}{r^4}\right).
\end{equation}
Averaging gives Eq.~\eqref{eq:SI_variance_relation}. Positive local
anisotropy and finite moments are assumed throughout.

Separate the fluctuation into a residual part and a
deformation-sensitive part,
\begin{equation}
\label{appB:eq:fluctuation_components}
\begin{aligned}
\delta r^{(\nu)}
&=\delta r_\infty^{(\nu)}+a^{(\nu)}(\zeta),
\\
\langle\delta r_\infty^{(\nu)}\rangle_\nu
&=\langle a^{(\nu)}\rangle_\nu=0,
\qquad
\langle\delta r_\infty^{(\nu)}a^{(\nu)}\rangle_\nu=0.
\end{aligned}
\end{equation}
The residual distribution is independent of $\zeta=I_1-3$.
Let $p(a;\zeta)$ denote the probability density of the
deformation-sensitive part. The probability flux in fluctuation space is $\mathcal Jp$, whereas $\mathcal J$ is the transport velocity. Conservation of probability gives
\begin{equation}
\label{eq:SI_probability_conservation}
\int_{\mathbb R}p(a;\zeta)\,{\rm d}a=1,
\qquad
\frac{\partial p}{\partial\zeta}
+
\frac{\partial}{\partial a}(\mathcal Jp)=0.
\end{equation}
Under the linear contraction
$\mathcal J=-ma/2$ in Eq.~\eqref{eq:SI_transport_velocity},
the trajectories and the transported density are
\begin{equation}
\label{eq:SI_transport_trajectory}
\begin{aligned}
a(\zeta)&=a(0)\exp(-m\zeta/2),
\\
p(a;\zeta)
&=\exp(m\zeta/2)\,
p_0\!\left(a\exp(m\zeta/2)\right).
\end{aligned}
\end{equation}
This scalar contraction preserves the zero mean and the initially
vanishing covariance in Eq.~\eqref{appB:eq:fluctuation_components}.

Define the transient variance by
\begin{equation}
\label{appB:eq:transient_variance}
\Delta_{\rm tr}(\zeta)
:=\langle(a^{(\nu)})^2\rangle_\nu
=\int_{\mathbb R}a^2p(a;\zeta)\,{\rm d}a.
\end{equation}
Assuming that the moment flux vanishes at the integration
boundaries, integration by parts yields
\begin{equation}
\label{eq:SI_variance_evolution}
\begin{aligned}
\frac{{\rm d}\Delta_{\rm tr}}{{\rm d}\zeta}
&=-\int_{\mathbb R}a^2
\frac{\partial}{\partial a}(\mathcal Jp)\,{\rm d}a
\\
&=2\int_{\mathbb R}a\mathcal Jp\,{\rm d}a
=-m\Delta_{\rm tr}.
\end{aligned}
\end{equation}
Consequently,
$\Delta_{\rm tr}(\zeta)=\Delta_{\rm tr}(0)\exp(-m\zeta)$.
With
\begin{equation}
\label{appB:eq:variance_components}
\Delta_\infty
:=
\langle(\delta r_\infty^{(\nu)})^2\rangle_\nu,
\qquad
\Delta_0=\Delta_\infty+\Delta_{\rm tr}(0),
\end{equation}
the total variance is Eq.~\eqref{eq:SI_variance_solution}.
Applying the same deformation-independent proportionality factor
in Eq.~\eqref{eq:SI_Warner_relation}, at fixed $r$, gives
Eq.~\eqref{eq:semisoft_law} to the order retained in the
weak-fluctuation expansion.
The exponential law therefore follows from the specified
contraction field together with probability conservation. Here
$\zeta$ labels the current deformation.

\subsection{Quadratic strain--director coupling}
\label{appB:quadratic_coupling}

Introduce the transverse projector
$\mathbf P_0=\mathbf I-\mathbf n_0\otimes\mathbf n_0$.
Contracting the TI stiffness tensor in Eq.~\eqref{eq:TI_stiffness}
gives the exact representation
\begin{equation}
\label{appB:eq:network_energy_projection}
\begin{aligned}
W_{\rm TI}^{\rm soft} = {}&C_1
\left(\mathbf n_0\cdot\mathbf E^{\rm e}\mathbf n_0\right)^2
+2C_4
\left(\mathbf P_0\mathbf E^{\rm e}\mathbf P_0\right)
:
\left(\mathbf P_0\mathbf E^{\rm e}\mathbf P_0\right)
+4C_5
\left\|\mathbf P_0\mathbf E^{\rm e}\mathbf n_0\right\|^2.
\end{aligned}
\end{equation}
For $C_1,C_4,C_5>0$, every independent component of the symmetric
tensor $\mathbf E^{\rm e}$ has a positive quadratic coefficient.
Thus the network energy vanishes precisely when
$\mathbf E^{\rm e}=\mathbf0$.

Expand about $\mathbf F=\mathbf I$ and $\mathbf n=\mathbf n_0$,
using the infinitesimal kinematics
\begin{equation}
\label{eq:SI_small_kinematics}
\mathbf F
\approx\mathbf I+\boldsymbol\varepsilon+\boldsymbol\Omega,
\qquad
\mathbf n
\approx\mathbf n_0+\boldsymbol\Omega^{\rm n}\mathbf n_0.
\end{equation}
Here $\boldsymbol\varepsilon$ is symmetric with
$\operatorname{tr}\boldsymbol\varepsilon=0$, while
$\boldsymbol\Omega$ and $\boldsymbol\Omega^{\rm n}$ are
skew-symmetric. The director rotation relative to the material is
\begin{equation}
\label{appB:eq:relative_director_rotation}
\widetilde{\mathbf n}
=(\boldsymbol\Omega^{\rm n}-\boldsymbol\Omega)\mathbf n_0,
\qquad
\widetilde{\mathbf n}\cdot\mathbf n_0=0.
\end{equation}
Substitution of the spontaneous stretches and
$\mathbf F^{\rm e}=(\mathbf V^{\rm s})^{-1}\mathbf F\mathbf U^{\rm s}$
gives, to first order,
\begin{equation}
\label{appB:eq:elastic_strain_linearization}
\begin{aligned}
&\mathbf n_0\cdot\mathbf E^{\rm e}\mathbf n_0
 \approx\mathbf n_0 \cdot \boldsymbol\varepsilon\mathbf n_0, \qquad
\mathbf P_0\mathbf E^{\rm e}\mathbf P_0
\approx\mathbf P_0\boldsymbol\varepsilon\mathbf P_0,
\\
&\mathbf P_0\mathbf E^{\rm e}\mathbf n_0
\approx
\frac{r+1}{2r^{1/2}}\,
\mathbf P_0\boldsymbol\varepsilon\mathbf n_0
-
\frac{r-1}{2r^{1/2}}\,\widetilde{\mathbf n}.
\end{aligned}
\end{equation}
Similarly, Eq.~\eqref{eq:semisoft_distortion} gives
\begin{equation}
\label{appB:eq:semisoft_distortion_linearization}
\mathbf n^{\rm F}
\approx\mathbf P_0\boldsymbol\varepsilon\mathbf n_0
+\widetilde{\mathbf n}.
\end{equation}
Both $I_1-3$ and $|\mathbf n^{\rm F}|^2$ are of second order near the incompressible reference state. Hence the deformation dependence of $D_{\rm ss}$ first contributes at fourth order in the energy, the quadratic energy involves only $D_{\rm ss}^{0}$.

Using these expansions in Eqs.~\eqref{eq:TI_soft_energy} and \eqref{eq:semisoft} yields the incompressible de Gennes-type energy \citep{de1980weak},
\begin{equation}
\label{eq:SI_deGennes_energy}
\begin{aligned}
W^{\rm linear}
={}&C_1
\left(\mathbf n_0\cdot\boldsymbol\varepsilon\mathbf n_0\right)^2
+2C_4
\left(\mathbf P_0\boldsymbol\varepsilon\mathbf P_0\right)
:
\left(\mathbf P_0\boldsymbol\varepsilon\mathbf P_0\right)
\\
&+4\bar C_5
\left\|\mathbf P_0\boldsymbol\varepsilon\mathbf n_0\right\|^2
+D_1\left\|\widetilde{\mathbf n}\right\|^2
-2D_2
\left(\mathbf P_0\boldsymbol\varepsilon\mathbf n_0\right)
\cdot\widetilde{\mathbf n},
\end{aligned}
\end{equation}
where
\begin{equation}
\label{eq:SI_linear_coefficients}
\begin{aligned}
\bar C_5
&=\frac{(r+1)^2}{4r}C_5+\frac{D_{\rm ss}^{0}}4,
\\
D_1&=\frac{(r-1)^2}{r}C_5+D_{\rm ss}^{0},
\qquad
D_2=\frac{r^2-1}{r}C_5-D_{\rm ss}^{0}.
\end{aligned}
\end{equation}

\subsection{Director relaxation and engineering moduli}
\label{appB:director_relaxation}

For $r>1$, $C_5>0$ and $D_{\rm ss}^{0}\ge0$, the coefficient $D_1$
is positive. Minimization of Eq.~\eqref{eq:SI_deGennes_energy} over
the relative director rotation gives
\begin{equation}
\label{appB:eq:relaxed_director}
\begin{aligned}
\widetilde{\mathbf n}
&=\frac{D_2}{D_1}\,
\mathbf P_0\boldsymbol\varepsilon\mathbf n_0.
\end{aligned}
\end{equation}
The relaxed energy is
\begin{equation}
\label{eq:SI_relaxed_energy}
\begin{aligned}
W^{\rm relaxed}
={}&C_1
\left(\mathbf n_0\cdot\boldsymbol\varepsilon\mathbf n_0\right)^2
+2C_4
\left(\mathbf P_0\boldsymbol\varepsilon\mathbf P_0\right)
:
\left(\mathbf P_0\boldsymbol\varepsilon\mathbf P_0\right)
+4C_5^{\rm eff}
\left\|\mathbf P_0\boldsymbol\varepsilon\mathbf n_0\right\|^2,
\end{aligned}
\end{equation}
where
\begin{equation}
\label{appB:eq:effective_shear_coefficient}
C_5^{\rm eff}
=\bar C_5-\frac{D_2^{\,2}}{4D_1}
=\frac{rC_5D_{\rm ss}^{0}}
{{C_5(r-1)^2}/{r}+D_{\rm ss}^{0}}.
\end{equation}
Thus $D_{\rm ss}^{0}>0$ restores a positive longitudinal shear stiffness, whereas $D_{\rm ss}^{0}=0$ gives the ideal soft limit $C_5^{\rm eff}=0$. This is a statement about the quadratic response at the reference state.

Choosing $\mathbf n_0=\mathbf e_3$, differentiation of Eq.~\eqref{eq:SI_relaxed_energy} gives the mixed incompressible Voigt relation
\begin{equation}
\label{eq:SI_relaxed_Voigt}
\begin{bmatrix}
\sigma_{11}\\
\sigma_{22}\\
\sigma_{33}\\
\sigma_{23}\\
\sigma_{13}\\
\sigma_{12}
\end{bmatrix}
=
-p
\begin{bmatrix}
1\\1\\1\\0\\0\\0
\end{bmatrix}
+
\begin{bmatrix}
4C_4&0&0&0&0&0\\
0&4C_4&0&0&0&0\\
0&0&2C_1&0&0&0\\
0&0&0&2C_5^{\rm eff}&0&0\\
0&0&0&0&2C_5^{\rm eff}&0\\
0&0&0&0&0&2C_4
\end{bmatrix}
\begin{bmatrix}
\varepsilon_{11}\\
\varepsilon_{22}\\
\varepsilon_{33}\\
2\varepsilon_{23}\\
2\varepsilon_{13}\\
2\varepsilon_{12}
\end{bmatrix}.
\end{equation}
Eliminating the pressure with the traction conditions and $\varepsilon_{11}+\varepsilon_{22}+\varepsilon_{33}=0$ gives the engineering moduli in Eq.~\eqref{eq:SI_apparent_moduli} \citep{ting1996anisotropic}. In particular, $E_\parallel=2(C_1+C_4)$, $G=2C_4$, and $E_\perp=4E_\parallel G/(E_\parallel+G)$.

For the reduced choice $C_1=2C_4$, $C_5=C_4$ and $\mu=2C_4$, Eq.~\eqref{appB:eq:network_energy_projection} reduces to $\mu\mathbf E^{\rm e}:\mathbf E^{\rm e}$. To quadratic order this is the incompressible expansion of $W_{\rm NC}^{\rm soft}$, so the TI and neo-classical models have the same initial response when their semisoft coefficients and anisotropy $r$ are also matched.

For oblique uniaxial stress in the $(\mathbf e_1,\mathbf e_3)$ plane, $\sigma_{22}=0$. For $D_{\rm ss}^{0}>0$, solving Eq.~\eqref{eq:SI_relaxed_Voigt} gives
\begin{equation}
\label{appB:eq:normal_and_shear_compliances}
\begin{aligned}
\varepsilon_{11}
&=\frac{\sigma_{11}}{E_\perp}
-\frac{\sigma_{33}}{2E_\parallel},
\\
\varepsilon_{33}
&=-\frac{\sigma_{11}}{2E_\parallel}
+\frac{\sigma_{33}}{E_\parallel},
\qquad
2\varepsilon_{13}=\frac{\sigma_{13}}{G_{\rm d}}.
\end{aligned}
\end{equation}
If the loading direction makes angle $\theta_0$ with $\mathbf n_0$, the stress components are
\begin{equation}
\label{appB:eq:oblique_stress}
\sigma_{11}=\sigma\sin^2\theta_0,
\qquad
\sigma_{33}=\sigma\cos^2\theta_0,
\qquad
\sigma_{13}=\sigma\sin\theta_0\cos\theta_0.
\end{equation}
The axial strain is therefore
\begin{equation}
\label{appB:eq:oblique_axial_strain}
\varepsilon_{11}\sin^2\theta_0
+2\varepsilon_{13}\sin\theta_0\cos\theta_0
+\varepsilon_{33}\cos^2\theta_0
=\frac{\sigma}{E(\theta_0)}.
\end{equation}
Substitution of Eqs.~\eqref{appB:eq:normal_and_shear_compliances} and \eqref{appB:eq:oblique_stress} recovers Eq.~\eqref{ch3:eq:Etheta}. The oblique modulus tends to zero in the ideal soft limit for $0<\theta_0<\pi/2$, while the two principal moduli remain finite.

\section{Parameter selection and fitting}
\label{app:parameter_identification}

This appendix specifies the parameters used in the controlled
model comparisons and in the experimental calibration.
The same symbols denote comparable quantities across models;
the model and calibration are identified by the table columns.

\subsection{Parameters for the controlled model comparison}
\label{appD:model_matching}

Table~\ref{tab:soft_elasticity_comparison_parameters} gives
the prescribed parameter sets for
Fig.~\ref{fig:soft_elasticity_comparison}.
The neo-classical and reduced TI references use the same $r$
and constant $D_{\rm ss}$, with $C_1=\mu$ and
$C_4=C_5=\mu/2$ for the reduced TI case.
Both references have $E_\parallel=E_\perp=12$ MPa.
The full TI example uses $E_\parallel\simeq12$ MPa and
$E_\perp\simeq4$ MPa, together with its listed semisoft
coefficient. The full TI variants in panel (d) retain $C_1=2C_4$
and the other reference parameters, while
$C_5/C_4$ takes the values $3$, $0.05$ and $+\infty$.
The reduced TI reference additionally satisfies $C_5=C_4$.

The common dimensionless symbol $a_{\rm ss}$ denotes
$2D_{\rm ss}/\mu$ in the neo-classical model and
$D_{\rm ss}/C_4$ in the TI model.
For the variable coefficient, $a_{\rm ss}^{0}=\alpha_0$
and $a_{\rm ss}^{\infty}=\alpha_\infty$.

\begin{table}[htbp]
\centering
\caption{Material parameters for
Fig.~\ref{fig:soft_elasticity_comparison}.}
\label{tab:soft_elasticity_comparison_parameters}
\begingroup
\normalfont\small
\setlength{\tabcolsep}{4pt}
\renewcommand{\arraystretch}{1.18}
\sisetup{detect-all,group-digits=false,table-number-alignment=center}
\begin{tabular*}{\linewidth}{@{\extracolsep{\fill}}ll *{5}{S[table-format=4.3]}@{}}
\toprule
\textbf{Parameter} & \textbf{Unit} &
\multicolumn{1}{c}{\shortstack{\textbf{Neo-}\\\textbf{classical}}} &
\multicolumn{1}{c}{\shortstack{\textbf{Reduced}\\\textbf{TI}}} &
\multicolumn{1}{c}{\shortstack{\textbf{Full TI}\\(a)--(c)}} &
\multicolumn{1}{c}{\shortstack{\textbf{TI 1}\\(d)}} &
\multicolumn{1}{c}{\shortstack{\textbf{TI 2}\\(d)}}\\
\midrule
$\mu$ & kPa & 4000 & \multicolumn{1}{c}{\textemdash} & \multicolumn{1}{c}{\textemdash} & \multicolumn{1}{c}{\textemdash} & \multicolumn{1}{c}{\textemdash}\\
$C_1$ & kPa & \multicolumn{1}{c}{\textemdash} & 4000 & 5455 & 4000 & 4000\\
$C_4$ & kPa & \multicolumn{1}{c}{\textemdash} & 2000 & 545.5 & 2000 & 2000\\
$C_5$ & kPa & \multicolumn{1}{c}{\textemdash} & 2000 & 545.5 & 6000 & 100\\
$D_{\rm ss}$ & kPa & 331.6 & 331.6 & 108.7 & 331.6 & 331.6\\
$a_{\rm ss}$ & 1 & 0.166 & 0.166 & 0.199 & 0.166 & 0.166\\
$r$ & 1 & 3 & 3 & 3 & 3 & 3\\
$C_1/C_4$ & 1 & \multicolumn{1}{c}{\textemdash} & 2 & 10 & 2 & 2\\
$C_5/C_4$ & 1 & \multicolumn{1}{c}{\textemdash} & 1 & 1 & 3 & 0.05\\
\bottomrule
\end{tabular*}
\par\smallskip
{\footnotesize\raggedright
Units are given in the second column; $1$ denotes a
dimensionless quantity. All cases use a constant semisoft modulus.
Values are rounded to at most four significant figures.\par}
\endgroup
\end{table}

\subsection{Calibration of the common TI parameter set}
\label{appD:initial_identification}

The experimental comparison in
Fig.~\ref{fig:stress-curves-initial-angles} uses the tensile
measurements of \citet{okamoto2021effect}.
The common TI set is anchored by
$E_\parallel=11.9$ MPa and $E_\perp=4.4$ MPa.
Equation~\eqref{eq:SI_apparent_moduli} then determines
$C_1=5.344$ MPa and $C_4=0.606$ MPa, to the displayed precision.
The common calibration additionally imposes $C_5=C_4$.
This is a parameter constraint: the initial angular modulus
constrains $G_{\rm d}$ through Eq.~\eqref{ch3:eq:Etheta},
but cannot separately determine $C_5$ and $D_{\rm ss}^{0}$.

With these network coefficients fixed, the remaining inputs
are $r$, $\alpha_0$, $\alpha_\infty$ and $m$.
These parameters were selected by manual adjustment against
the experimental stress curves. The perpendicular rotation
onset, the level and shape of the rotating response, and the
subsequent tensile hardening guide the joint adjustment.
The onset samples the current coefficient
$D_{\rm ss}(I_{1c})$ in Eq.~\eqref{ch3:eq:onset}, so it
constrains a combination of the three semisoft parameters.
The coefficient $\alpha_0$ sets the initial semisoft resistance,
$\alpha_\infty$ its residual level, and $m$ the deformation
scale of its change, as specified by Eq.~\eqref{eq:semisoft_law}.
The reported values are listed in
Table~\ref{appD:tab:experimental_parameters}.

This one parameter set is used for the $0^\circ$, $35^\circ$,
$57^\circ$ and $90^\circ$ stress curves labeled
\emph{uniform}. The initial angular-modulus curve is evaluated
from the same inputs.
The agreement across orientations is assessed with all
material parameters held fixed.

\subsection{Neo-classical reference and individual-angle fits}
\label{appD:comparison_protocol}

For the neo-classical reference,
$\mu=4.045$ MPa and $r=6.9$ are used across orientations.
Its constant semisoft coefficient is $\alpha_0=0.095$ and $D_{\rm ss}=0.192$ MPa.
The resulting principal moduli are both $3\mu$, as reported
in Table~\ref{appD:tab:experimental_parameters}.

The \emph{independent} curves were obtained by manual
recalibration for $57^\circ$ and $35^\circ$ separately.
Both retain the common $C_1,C_4$, and hence the same principal
Young's moduli, while the listed shear, anisotropy and semisoft
parameters are adjusted against the respective stress curve. The two individual sets share the listed
$r$, which differs from the common-set value.
They apply only to their indicated angle.

\begin{table}[htbp]
\centering
\caption{Material parameters and initial moduli for the
experimental comparison in
Fig.~\ref{fig:stress-curves-initial-angles}.}
\label{appD:tab:experimental_parameters}
\begingroup
\normalfont\small
\setlength{\tabcolsep}{4pt}
\renewcommand{\arraystretch}{1.18}
\sisetup{detect-all,group-digits=false,table-number-alignment=center}
\begin{tabular*}{\linewidth}{@{\extracolsep{\fill}}ll *{4}{S[table-format=2.4]}@{}}
\toprule
& & \multicolumn{2}{c}{\textbf{Common across angles}} &
\multicolumn{2}{c}{\textbf{Individual TI}}\\
\cmidrule(lr){3-4}\cmidrule(lr){5-6}
\textbf{Parameter} & \textbf{Unit} &
\multicolumn{1}{c}{\shortstack{\textbf{Neo-}\\\textbf{classical}}} &
\multicolumn{1}{c}{\shortstack{\textbf{Common}\\\textbf{TI}}} &
\multicolumn{1}{c}{$57^\circ$} &
\multicolumn{1}{c}{$35^\circ$}\\
\midrule
$\mu$ & MPa & 4.045 & \multicolumn{1}{c}{\textemdash} & \multicolumn{1}{c}{\textemdash} & \multicolumn{1}{c}{\textemdash}\\
$C_1$ & MPa & \multicolumn{1}{c}{\textemdash} & 5.344 & 5.344 & 5.344\\
$C_4$ & MPa & \multicolumn{1}{c}{\textemdash} & 0.606 & 0.606 & 0.606\\
$C_5$ & MPa & \multicolumn{1}{c}{\textemdash} & 0.606 & 0.8201 & 0.4775\\
$r$ & 1 & 6.9 & 6.95 & 5.986 & 5.986\\
$D_{\rm ss}$ & MPa & 0.192 & \multicolumn{1}{c}{\textemdash} & \multicolumn{1}{c}{\textemdash} & \multicolumn{1}{c}{\textemdash}\\
$\alpha_0$ & 1 & \multicolumn{1}{c}{0.095} & 0.66 & 1.152 & 0.7451\\
$\alpha_\infty$ & 1 & \multicolumn{1}{c}{\textemdash} & 0.3511 & 0.4944 & 0.3334\\
$m$ & 1 & \multicolumn{1}{c}{\textemdash} & 0.3111 & 0.4607 & 0.5712\\
\midrule
$E_\parallel$ & MPa & 12.14 & 11.9 & 11.9 & 11.9\\
$E_\perp$ & MPa & 12.14 & 4.4 & 4.4 & 4.4\\
$G_{\rm d}$ & MPa & 0.5154 & 0.9662 & 1.671 & 1.06\\
\bottomrule
\end{tabular*}
\par\smallskip
{\footnotesize\raggedright
Dimensional quantities are in MPa; $1$ denotes a dimensionless
quantity. Common parameter sets apply
across orientations; individual TI sets apply only to the
indicated angle. Initial moduli below the internal rule are
derived from the listed parameters. Values are rounded to at
most four significant figures.\par}
\endgroup
\end{table}

In this table, $\alpha_\infty=D_{\rm ss}^{\infty}/C_4$
is the residual coefficient; some labels in
Fig.~\ref{fig:semisoft_decay_comparison} instead give the
ratio $\alpha_\infty/\alpha_0$.
The listed $E_\parallel,E_\perp,G_{\rm d}$ are calculated
from each parameter set, rather than fitted as additional
independent parameters. Numerical calculations retain the
unrounded inputs; the tables report at most four significant
figures.

The individual fits quantify the improvement obtainable by
angle-specific recalibration. The common set provides the
assessment of a single constitutive description across angles.
The stress curves and initial moduli constrain several parameter
combinations, so the reported values represent one calibration;
director and lateral-deformation data would add independent
constraints.


\bibliographystyle{elsarticle-harv}
\bibliography{references}


\end{document}